\documentclass{amsart}
\usepackage[english]{babel}
\usepackage{amsmath}

\makeatletter
\renewcommand{\@eqnnum}{\hb@xt@.01\textwidth{\hfil\theequation}}
\makeatletter

\usepackage{graphicx}
\usepackage[foot]{amsaddr}
\usepackage{xcolor}
\usepackage{subcaption}
\usepackage{float}
\usepackage{multirow}
\newcommand{\fig}[1]{Fig.~\ref{#1}}{\color{blue}}

\usepackage{comment}
\usepackage{stackengine,graphicx}
\usepackage{tikz}
\usepackage[top=1in, bottom=1.25in, left=1.25in, right=1.25in]{geometry}
\usepackage{overpic}

\usepackage{hyperref}
    \hypersetup{
        colorlinks   = true,
        allcolors   =blue    
    }
\usepackage{tikz}
\usetikzlibrary{calc,fit,backgrounds}
\usetikzlibrary{shapes.geometric, arrows}
\usetikzlibrary{chains,
                fit,
                positioning,
                shapes}

\def\u{{\bm u}}

\def\U{{\bm U}}

\def\0{\boldsymbol{0}}

\newcommand{\bm}[1]{\mbox{\boldmath{$#1$}}}

\begin{document}

\title[A comparison of data-driven ROMs for atmospheric flow]{A comparison of data-driven Reduced Order Models for the simulation of mesoscale atmospheric flow}
\author{Arash Hajisharifi$^1$, Michele Girfoglio$^1$, Annalisa Quaini$^{2,*}$, and Gianluigi Rozza$^1$}
\address{$^1$ mathLab, Mathematics Area, SISSA, via Bonomea 265, I-34136 Trieste, Italy
}
\address{$^2$ Department of Mathematics, University of Houston, Houston TX 77204, USA}
\address{$^*$ Corresponding author: quaini@math.uh.edu}

\begin{abstract}
The simulation of atmospheric flows by means of traditional discretization methods remains computationally intensive, hindering the achievement of high forecasting accuracy in short time frames.
In this paper, we apply three reduced order models that have successfully reduced the computational time for different applications in computational fluid dynamics while preserving accuracy: Dynamic Mode Decomposition (DMD), Hankel Dynamic Mode Decomposition (HDMD), and Proper Orthogonal Decomposition with Interpolation (PODI). The three methods are compared in terms of computational time and accuracy in the simulation of two well-known benchmarks for mesoscale flow. The accuracy of the DMD and HDMD solutions deteriorates rather quickly as the forecast time window expands, although these methods are designed to predict the dynamics of a system. The reason is likely the strong nonlinearity in the benchmark flows.
The PODI solution is accurate for the entire duration of the time interval of interest thanks to the use of interpolation with radial basis functions. This holds true also when the model features a physical parameter expected to vary in a given range, as is typically the case in weather prediction. 
\end{abstract}

\maketitle


\section{Introduction}




Despite a continuous increase in  computational power, simulations of atmospheric flow using classical discretization methods
(e.g., finite element methods or finite volume methods) remain computationally expensive. Given the large number of simulations required to quantify uncertainty in weather prediction, alternatives to such discretization methods, also called Full Order Models (FOMs), are needed to reduce the computational time and allow for improved prediction accuracy in short time frames. 

For over a couple of decades, Reduced Order Models (ROMs) have emerged a methodology of choice to reduce the computational burden when FOM simulations have to be carried out for several (physical) parameter values, as in the case of uncertainty quantification, or for long periods of time, as in the case of forecasts. ROMs replace the FOM of choice with a lower-dimensional approximation that captures the essential behavior of the system.  
This is achieved through a two-step procedure.
In the first step, called \emph{offline
phase}, one constructs a database of several FOM solutions associated to given times and/or
physical parameter values. An example of  a physical parameter for an atmospheric flow problem
could be an initial temperature perturbation 
with magnitude expected to vary in a given range.
The database of FOM solutions is used to generate a reduced basis, which is (hopefully much) smaller than the high-dimensional FOM basis but still preserves the essential features of the system. 
In the second step, called \emph{online phase}, one uses this reduced basis to quickly compute the solution for newly specified times and/or parameter values. Note that, while the offline phase is performed once and for all, the online phase is performed as many times as needed. 
For a comprehensive review on ROMs, the reader is referred to, e.g., \cite{peter2021modelvol1,benner2020modelvol2,benner2020modelvol3,bochev2009least,hesthaven2016certified, malik2017reduced,rozza2008reduced}.

The ROMs that have been successfully applied to fluid dynamics problems could be divided into two major categories: projection-based vs.~data-driven. In general terms, projection-based ROMs project the governing equations onto the low-dimensional subspace spanned by the basis functions. In order to implement a projection-based ROM efficiently, one needs access to the source code of the FOM solver. 
On the other hand, data-driven ROMs rely on available data from the high-dimensional FOM system to directly learn a reduced order model without explicitly considering the underlying equations. While projection-based methods aim to preserve the governing equations of the high-dimensional system in the reduced model, data-driven methods construct the reduced order model purely from the data, i.e., they learn the relationships and patterns observed in the data. 
Hence, a data-driven ROM is blind to the mathematical model and treats the FOM solver as a black box. For this reason, data-driven ROMs are also called non-intrusive. 

In this paper, we focus on data-driven ROMs. 
The efficiency of projection-based ROMs
is rather limited in the case of nonlinear problems as these problems often require hyper-reduction techniques (see, e.g., \cite{barrault2004empirical, chaturantabut2010nonlinear})
that are problem-dependent
and computationally expensive. So, non-intrusive ROMs are to be preferred for applications where a high speed-up is required. Since atmospheric flows are highly nonlinear and high speed-up is desirable for forecasts, the aim of this paper is to compare in terms of accuracy and computational time
three data driven ROMs that have been successfully applied to different fluid dynamics problems: Dynamic Mode Decomposition (DMD), Hankel Dynamic Mode Decomposition (HDMD) and Proper Orthogonal Decomposition with Interpolation (PODI). We start from DMD because it is specifically designed to predict the future behavior of a system \cite{kutz2016dynamic,schmid2010dynamic, schmid2011applications, tu2013dynamic} and 
it was shown to work well for problems like axisymmetric jet flow \cite{schmid2011application}, annular liquid sheets \cite{duke2012experimental}, and turbulent cavity flow \cite{seena2011dynamic}. 
The natural next choice is HDMD because it improves the DMD algorithm with time-delay embedding \cite{arbabi2017ergodic,  curtis2023machine, fujii2019data, jiang2015study,vasconcelos2019dynamic, yang2021synchronized}. The result is that HDMD can predict more accurately and for longer periods of time systems exhibiting strong nonlinear dynamics \cite{frame2022space,vasconcelos2019dynamic}. {HDMD has been successfully applied to simulate, e.g., periodic cavity flow \cite{arbabi2017ergodic,Hess2023}, electromechanical systems  \cite{yang2021synchronized}, 
and biological systems \cite{fujii2019data}.}
PODI differs from DMD and HDMD in that it is not designed to forecast the system evolution, but rather to interpolate solutions in a parameter space, where time is one of possibly many parameters of interest. 
So far, PODI has been applied to perform parametric studies for problems stemming from, e.g., thermo-mechanics \cite{shah2022finite}, hemodynamics \cite{girfoglio2022non}, chemical \cite{hajisharifi2023non} and naval \cite{demo2018efficient, demo1} engineering, 
and aeronautics \cite{ripepi2018reduced}.

POD, which lies at the core of many ROMs, is often referred to as Empirical Orthogonal Function (EOF) analysis in the geophysical fluid dynamics community. It has been applied to reanalysis data to identify spatio-temporal coherent meteorological patterns and teleconnections, e.g.,  the Madden-Julian Oscillation, the Quasi-Biennial Oscillation, and the El Nino-Southern Oscillation. See, e.g., \cite{lario2022neural,pawar2022equation,schmidt2019spectral}. 
These phenomena involve large spatial and temporal scales: they result from the interaction of global circulation effects and happen over periods of time ranging from several months to many years. The EOF analysis uses data on the global scale and considers time as the only parameter. In addition, it is mostly limited to system identification. Very recently, a data-driven ROM based on EOF analysis has been used for pattern prediction, specifically to forecast the weekly average sea surface temperature
\cite{pawar2022equation}.  
Other data-driven methods borrowed from 
Machine Learning have been applied to global weather forecasting. See, e.g., \cite{pathak2018model,rasp2021data,  schultz2021can,weyn2019can}. However, these methods cannot be strictly categorized as reduced order modeling since no reduced basis is generated. 


There work presented in this paper distinguishes itself from the EOF analyses in the literature in a few aspects. The main novelty lies in the fact that we apply ROMs for both system identification and forecast of regional atmospheric flows.  
We focus on a spatial scale of a few kilometers and a time scale of a few hours, not on the global circulation for long periods of time, with an obvious difference in resolution. Finally, with PODI we perform a parametric study that includes a physical parameter as well, i.e., our analysis is not limited to time as the only parameter.

This paper is organized as follows. Sec.~\ref{sec:FOM} describes the compressible Euler equations for low Mach stratified flows and gives details about the full order model. Sec. \ref{sec:ROM} presents the main ingredients of the three ROMs under consideration. Sec.~\ref{sec:res} reports the comparison of the three ROMs using two well-known 2D benchmark problems involving stratified and gravity driven atmospheres: the rising thermal bubble 
and the density current. 
Finally, Sec.~\ref{sec:conc} provides conclusions and perspectives.

\section{The full order model}\label{sec:FOM}

We consider the dynamics of dry atmosphere, i.e., we neglect the effects of moisture. In addition, we neglect solar radiation and heat flux from the ground. 
We assume that dry air behaves like an ideal gas. Then, the equations describing the mass, momentum, and energy conservation in a spatial domain of interest $\Omega$ over a certain time interval $(0, t_f]$ are given by   
\begin{align}
&\frac{\partial \rho}{\partial t} + \nabla \cdot ( \rho \u ) = 0 &&\text{in} \ \Omega \ \times \ (0, t_f], \label{continuity} \\
&\frac{\partial (\rho \u)}{\partial t} + \nabla \cdot ( \rho \u \otimes \u) + \nabla p + \rho g \widehat{\mathbf k} = \boldsymbol{0}  &&\text{in} \ \Omega \ \times  \ (0, t_f], \label{momentu} \\
&\frac{\partial (\rho e)}{\partial t} + \nabla \cdot ( \rho \u e) + \nabla \cdot (p \u) = 0  &&\text{in} \ \Omega \ \times  \ (0, t_f],
\label{energy}
\end{align}
where $\rho$ is the air density, $\u$ is the wind velocity, $p$ is the pressure, $\widehat{\mathbf k}$ is the unit vector aligned with the vertical axis $z$, $g$ is the gravitational constant, and $e$ is the total energy density. Note that 
$e = c_v T + |\u|^2/2 + g z$ where $c_{v}$ is the specific heat capacity at constant volume, $T$ is the absolute temperature,  and $z$ is the vertical coordinate. To close system \eqref{continuity}-\eqref{energy}, 
we have the thermodynamics equation of state for ideal gases:
\begin{align}
p = \rho R T,
\label{EQ_of_State}
\end{align}
where $R$ is the specific gas constant of dry air. 

Introducing the following splitting of the pressure
\begin{align}
p = p^\prime + \rho g z,
\label{EQ_p_splitting}
\end{align}
where $\rho g z$ is a background state and
$p^\prime$ is a fluctuation with respect to it, eq.~\eqref{momentu} can be recast as:
\begin{align}
\frac{\partial (\rho \u)}{\partial t} + \nabla \cdot ( \rho \u \otimes \u) + \nabla p^\prime +  g z \nabla \rho = 0  \hspace{3cm} \text{in} \ \Omega \ \times  \ (0, t_f].
\label{momentu2}
\end{align}

Let $c_p$ be the specific heat capacity at constant pressure for dry air.
By introducing the specific enthalpy $h = c_v T + p/ \rho  = c_p T$, the total energy can be written as $e = h - {p}/{\rho} + K + gz$, 
where $K = |\u|^2/2$ is the kinetic energy density. Then, eq. \eqref{energy} can be rewritten as:
\begin{align}
\frac{\partial (\rho h)}{\partial t} + \nabla \cdot ( \rho \u h) + \frac{\partial (\rho K)}{\partial t} + \nabla \cdot (\rho \u K) -\frac{\partial p}{\partial t} + \rho g \u \cdot \widehat{\mathbf k}= 0  \hspace{1cm} \text{in} \ \Omega \ \times  \ (0, t_f].
\label{energy2}
\end{align}
To obtain \eqref{energy2}, we have also employed eq. \eqref{continuity}. 

Since problem \eqref{continuity},\eqref{EQ_of_State}-\eqref{energy2} does not feature any dissipation mechanism, 
perturbations due to numerical error can lead to a simulation breakdown or large, unphysical oscillations in the computed solution. Hence, numerical stabilization is needed. Typically, it amounts to introducing introducing additional (dissipative) terms in eq.~\eqref{momentu2} and \eqref{energy2} as follows: 
\begin{align}
&\frac{\partial (\rho \u)}{\partial t} +  \nabla \cdot (\rho \u \otimes \u) + \nabla p' + gz \nabla \rho -  \nabla \cdot (2 \mu_a \boldsymbol{\epsilon}(\u)) + \nabla \left(\frac{2}{3}\mu_a \nabla \cdot \u \right)= 0 &&\text{in } \Omega \times (0,t_f],  \label{eq:mom_LES} \\
&\frac{\partial (\rho h)}{\partial t} +  \nabla \cdot (\rho \u h) + 
\frac{\partial (\rho K)}{\partial t} +  \nabla \cdot (\rho \u K) - \dfrac{\partial p}{\partial t}  +  
\rho g \u \cdot \widehat{\mathbf k}  - \nabla \cdot \left(\frac{\mu_a}{Pr} \nabla h \right) = 0 &&\text{in } \Omega \times (0,t_f],
\label{eq:ent_LES}
\end{align}
where $\mu_a$ is an artificial viscosity, $\boldsymbol{\epsilon}(\u) = (\nabla \u + (\nabla \u)^T)/2$ is the strain-rate tensor, and $Pr$ is the Prandtl number, i.e., the dimensionless number defined as the ratio of momentum diffusivity to thermal diffusivity. Typically, the introduction of artificial viscosity $\mu_a$ serves the dual purpose of achieving stabilization and Large Eddy Simulation (LES). See, e.g., \cite{clinco2023filter, girfoglio2023validationAIP, marras2015stabilized}. 

The full order model in this paper is given by the stabilized Euler equations in the formulation \eqref{continuity},\eqref{EQ_of_State},\eqref{EQ_p_splitting},\eqref{eq:mom_LES},\eqref{eq:ent_LES}.

A quantity of interest for atmospheric problems is the potential temperature $\theta$. Many authors choose to formulate the Euler equations with $\theta$ as a variable. Instead, we compute it from $T$ and $p$ using the following definition:
\begin{align}
\theta = \frac{T}{\pi},  \quad \pi = \left(\frac{p}{p_0}\right)^{\frac{R}{c_p}},
\label{pot_temp}
\end{align}
where $p_0 = 10^5$ Pa, which is the atmospheric pressure at the ground.
Let us also define the potential temperature fluctuation $\theta^ \prime$, which is the difference between $\theta$ and its typical hydrostatic value $\theta_0$: 
\begin{align}
\theta ^ \prime (x,y,z,t) = \theta (x,y,z,t) - \theta_0(z).
\label{pot_temp2}
\end{align}
See, e.g., \cite{kelly2012continuous} for more details.

\subsection{Some details about the full order method (FOM)}

If one is not careful in designing an efficient numerical scheme, a solver for the full order model \eqref{continuity},\eqref{EQ_of_State},\eqref{EQ_p_splitting},\eqref{eq:mom_LES},\eqref{eq:ent_LES} could be computationally intensive. In order to contain the computational cost, we use a splitting scheme thoroughly described in \cite{girfoglio2023validationAIP}.
Here, we report only some details. 

To discretize in time, we introduce a time step
$\Delta t \in \mathbb{R}$ to partition time interval $(0,t_f]$ and obtain time levels $t^n = t_0 + n \Delta t$, with $n = 0, ..., N_{tf}$ and $t_f = 0 + N_{tf} \Delta t$.
For the discretization of the Eulerian time derivatives in \eqref{continuity}, \eqref{eq:mom_LES}, and \eqref{eq:ent_LES}, we adopt the Backward Euler scheme. 
In eq.~\eqref{eq:mom_LES} and \eqref{eq:ent_LES}, the treatment of the convective terms is semi-implicit while the treatment of the diffusive terms is implicit. On the other hand, eq.~\eqref{continuity} is treated explicitly. 
For the space discretization, 
the computational domain $\Omega$ is partitioned into cells or control volumes $\Omega_i$,
with $i = 1, \dots, N_{c}$, where $N_{c}$ is the total number of cells in the mesh. 
We adopt second-order finite volume schemes. 
Finally, in order to decouple the computation of the pressure from the computation of the velocity we use the PISO algorithm \cite{patankar1983calculation,issa1986solution,moukalled2016finite}.

This FOM is implemented within GEA (Geophysical and Environmental Applications) \cite{GEA,GQR_GEA, clinco2023filter, girfoglio2023validationAIP}, an open-source package for atmosphere and ocean modeling based on the finite volume C++ library OpenFOAM\textsuperscript{\textregistered}.

We would like to point out that, although we have
made specific choices for the full order method (e.g., spatial discretization via a finite volume method), the conclusions that we will draw about
the reduce order methods presented in the next section are expected to hold also for different 
full order methods (e.g., finite element methods).

\section{The reduced order model}\label{sec:ROM}


Let us assume that the PDE problem 
described in Sec.~\ref{sec:FOM} depends on some physical parameters that vary over a certain interval. 
Let $d$ be the number of parameters of interest and $\bm{\pi}$ the vector that stores them. In addition, 
let $\mathcal{P} \subset \mathbb{R}^d$ be parameter space with $\bm{\pi} \in \mathcal{P}$. 
Although the time $t$ could be treated as a parameter,  
we do not store it in
$\bm{\pi}$ and deal with it separately. 

The basic assumption of ROM for a PDE problem depending on time $t$ and parameter vector $\bm{\pi}$ is that any solution can be represented in terms of a linear combination of a reduced number of global basis functions, that depend exclusively on space $\bm{x}$, 
with the weights of the linear combination depending only on $t$ and $\bm{\pi}$.
In the case of the potential temperature perturbation $\theta^\prime$, which is our variable of interest, this is written as:
\begin{align}
   \theta^\prime(\bm{x}; t, \bm{\pi})  \approx \theta^\prime_r (\bm{x}; t, \bm{\pi})= \sum_{i=1}^{N_{\theta^\prime}} {\alpha}_i(t, \bm{\pi}) {\varphi}_i(\bm{x}),
   \label{eq:ROM_1}
\end{align}
where $\theta^\prime_r$ is the reduced order approximation of $\theta^\prime$, $N_{\theta^\prime}$ is the number of basis functions, the ${\varphi}_i$ are the basis functions and the ${\alpha}_i$ are the weights of the linear combination. If the time $t$ is the only parameter of interest, then eq.~\eqref{eq:ROM_1} becomes
\begin{align}
\theta^\prime(\bm{x}; t)  \approx \theta^\prime_r (\bm{x}; t) = \sum_{i=1}^{N_{\theta^{\prime}}}  \mathbb{{\alpha}}_i(t)\mathbb{{\varphi}}_i(\bm{x}).
\label{eq:ROM_approx_t}
\end{align}
While we focus on approximating $\theta^\prime$, approximations similar to \eqref{eq:ROM_1} and \eqref{eq:ROM_approx_t} can be applied to any other variable of interest, either primal (i.e., unknown of the original problem, density, velocity, pressure and the specific enthalpy) or derived (like $\theta^\prime$ itself).

In this paper, we focus on three data-driven ROMs that have not been applied to mesoscale atmospheric flows yet: Dynamic Mode Decomposition (DMD), Hankel DMD (HDMD), and
Proper Orthogonal Decomposition with Interpolation (PODI).  All these methods rely on the Singular Value Decomposition (SVD) algorithm as a main tool to compute the basis functions $\varphi_i$. DMD and Hankel DMD are mainly designed to predict the future behavior of a system, so they are intrinsically suited for problems where the time is the parameter of interest. On the other hand, PODI relies primarily on interpolation procedures in the parameter space, so it is not designed to forecast the time evolution of the system but it interpolates solutions dependent on time and parameter $\bm \pi$ alike.

In the three subsections below, we report the main ingredients of the DMD and HDMD algorithms, and the PODI approach.

\subsection{Dynamic Mode Decomposition}\label{sec:DMD}

Introduced in \cite{schmid2010dynamic},
DMD is a useful tool to effectively extract the dominant dynamic flow structure from a unsteady flow field
\cite{schmid2011application, schmid2011applications,tezzele2018model,rowley2009spectral}.
In particular, DMD forecasts future states of a non-linear time-dependent system through a linear combination of few main structures evolving linearly. Such a feature makes the DMD appealing for atmospheric problems and weather forecasts.
In this work, we will adopt DMD to reconstruct, and more importantly, forecast the time evolution of $\theta^{\prime}$ through \eqref{eq:ROM_approx_t}. 

In this section, we briefly present the main ingredients of the DMD algorithm concerning the computation of the basis functions $\varphi_i$ and of the weights $\mathbb{{\alpha}}_i$ in \eqref{eq:ROM_approx_t}.
For more details, we refer the reader to, e.g., \cite{kutz2016dynamic,tu2013dynamic,andreuzzi2021dynamic}. 

Recall that $N_c$ denotes the number of degrees of freedom of the full order solution.
Let $\theta_h^{\prime} (\bm{x}; t^i) \in \mathbb{R}^{N_c}$, with $i = 1, \dots, N_t$, be the full order solution (also called snapshot) computed at  time instant $t^i$. 
As mentioned above, DMD is designed to predict the future of a system, thus $N_t<N_{tf}$.
The key idea of DMD is that there exists a governing operator matrix, denoted as $\mathbb{\bm{A}}$,  that maps $\theta^{\prime}_h(\bm{x},t^{i})$ to $\theta^{\prime}_h(\bm{x},t^{i+1})$ and approximates the nonlinear dynamics of the system, i.e.:
\begin{align}
    \theta^{\prime}_h(\bm{x},t^{i+1}) = \mathbb{\bm{A}} \theta^{\prime}_h(\bm{x},t^{i}). \label{eq:DMD1}
\end{align}
To find $\mathbb{\bm{A}} \in \mathbb{R}^{N_c \times N_c}$, we build two snapshot matrices in $\mathbb{R}^{N_c \times \left(N_t-1\right)}$: 
\begin{align}
\mathbb{\bm{S}}_1 = [\theta^{\prime}_h(\bm{x}, t^1), \dots, \theta^{\prime}_h(\bm{x}, t^{N_t - 1})] \quad \text{and} \quad \mathbb{\bm{S}}_2 = [\theta^{\prime}_h(\bm{x}, t^2), \dots, \theta^{\prime}_h(\bm{x}, t^{N_t})].   \label{eq:S1S2_DMD}
\end{align}
Notice that $\mathbb{\bm{S}}_1$ contains the first
$N_t-1$ snapshots, while $\mathbb{\bm{S}}_2$ contains the last $N_t-1$ snapshots. By rewriting \eqref{eq:DMD1} in matrix form, we obtain:
\begin{align}
    \mathbb{\bm{S}}_2 = \mathbb{\bm{A}} \mathbb{\bm{S}}_1. \label{eq:DMD2A}
\end{align}
Then, we perform the SVD of the matrix $\mathbb{\bm{S}}_1$:
\begin{align}
\mathbb{\bm{S}}_1  = \mathbb{\bm{U}} \mathbb{\bm{\Sigma}}\mathbb{\bm{V}}^T, 
\label{eq:DMD11}
\end{align}
where $\mathbb{\bm{U}} \in \mathbb R^{{N_c}\times N_c}$ is the orthogonal matrix whose columns are the left singular vectors, $\bm{\Sigma} \in \mathbb R ^{N_c\times \left(N_t - 1\right)}$ is the rectangular diagonal matrix containing the singular values $\sigma_1 \geq \sigma_2 \geq ... \geq \sigma_{\min\{N_c, N_t - 1\}} \geq 0$, and $\bm{V} \in 
\mathbb{R}^{\left(N_t - 1\right) \times \left(N_t - 1\right)}$ is the orthogonal matrix whose columns are the right singular vectors. Symbol $^T$  denotes the conjugate transpose.

In order 
to form a basis for the reduced space, we adopt the POD algorithm \cite{shah2022finite, hajisharifi2023non, demo2018efficient,demo1, ripepi2018reduced}. 
Let $R \leq \min\{N_c, N_t - 1\}$ be the rank of the matrix $\mathbb{\bm{S}}_1$. The POD space is spanned by the first $R$ columns of the matrix $\mathbb{\bm{U}}$. Then, the reduced space is constructed by retaining the first $N_{\theta'} < R$ columns, called POD modes. 
The value of $N_{\theta'}$ is commonly chosen to reach a user-provided threshold $\delta$ for the cumulative
energy of the singular values:
\begin{align}
E = \frac{\sum_{i=1}^{N_{\theta^{\prime}}}\sigma_i}{\sum_{i=1}^{R}\sigma_i} \ge \delta.
\label{k_equation}
\end{align}
Once we have $N_{\theta'}$, we can introduce $\mathbb{\bm{U}}_{N_{\theta^{\prime}}} \in \mathbb R^{{N_c}\times {N_{\theta^{\prime}}} }$, which comes from retaining the first $N_{\theta'}$ columns of $\mathbb{\bm{U}}$, 
$\bm{\Sigma}_{N_{\theta^{\prime}}} \in \mathbb R ^{{{N_{\theta^{\prime}}}}\times {N_{\theta^{\prime}}}}$, which comes from keeping 
the first $N_{\theta'}$ columns and rows of $\bm{\Sigma}$, and $\bm{V}_{N_{\theta^{\prime}}} \in \mathbb{R}^{ \left(N_t-1\right) \times N_{\theta^{\prime}}}$, which is obtained from the first $N_{\theta'}$ columns of $\bm{V}$.
Then, matrix $\mathbb{\bm{S}}_1$ is approximated as follows: 
\begin{align}
\mathbb{\bm{S}}_1 \approx \mathbb{\bm{U}}_{N_{\theta'}} \mathbb{\bm{\Sigma}}_{N_{\theta'}}\mathbb{\bm{V}}_{N_{\theta'}}^T.
\label{eq:DMD_new}
\end{align}

By plugging \eqref{eq:DMD_new} into \eqref{eq:DMD2A}, $\mathbb{\bm{A}}$ can be approximated as: 
\begin{align}
\mathbb{\bm{A}}  \approx  \mathbb{\bm{S}}_2 \mathbb{\bm{V}}_{N_{\theta^{\prime}}} \bm{\Sigma}_{N_{\theta^{\prime}}}^{-1} \mathbb{\bm{U}}_{N_{\theta^{\prime}}}^T.
\end{align}
This approximation is projected onto the POD modes to get the reduced operator matrix
$\bm{A}_{N_{\theta^{\prime}}}\in \mathbb R ^{{{N_{\theta^{\prime}}}}\times {N_{\theta^{\prime}}}}$:
\begin{align}
 \mathbb{\bm{A}}_{N_{\theta^{\prime}}} = \mathbb{\bm{U}}_{N_{\theta^{\prime}}}^T \mathbb{\bm{A}} \mathbb{\bm{U}}_{N_{\theta^{\prime}}} =  \mathbb{\bm{U}}_{N_{\theta^{\prime}}}^T \mathbb{\bm{S}}_2 \mathbb{\bm{V}}_{N_{\theta^{\prime}}} \bm{\Sigma}_{N_{\theta^{\prime}}}^{-1} \mathbb{\bm{U}}_{N_{\theta^{\prime}}}^T \mathbb{\bm{U}}_{N_{\theta^{\prime}}} = \mathbb{\bm{U}}_{N_{\theta^{\prime}}}^T \mathbb{\bm{S}}_2 \mathbb{\bm{V}}_{N_{\theta^{\prime}}} \bm{\Sigma}_{N_{\theta^{\prime}}}^{-1}.
\label{eq:DMD2}
\end{align}

Next, we perform the eigendecomposition of $\mathbb{\bm{A}}_{N_{\theta^{\prime}}}$, i.e. we solve the following eigenvalue problem:
\begin{align}
\mathbb{\bm{A}}_{N_{\theta^{\prime}}} \mathbb{\bm{W}} = \mathbb{\bm{W}} \mathbb{\bm{\Lambda}},
\label{eq:DMD3}
\end{align}
where $\mathbb{\bm{\Lambda}} \in \mathbb R ^{{{N_{\theta^{\prime}}}}\times {N_{\theta^{\prime}}}} $ is the diagonal matrix containing the eigenvalues $\lambda_1 \geq \lambda_2 \geq ... \geq \lambda_{N_{\theta'}} \geq 0$ of $\mathbb{\bm{A}}_{N_{\theta^{\prime}}}$ and $\mathbb{\bm{W}} \in \mathbb R ^{{{N_{\theta^{\prime}}}}\times {N_{\theta^{\prime}}}} $ is the matrix whose columns are the eigenvectors of $\mathbb{\bm{A}}_{N_{\theta^{\prime}}}$.
The eigenvalues of $\mathbb{\bm{A}}_{{N_{\theta'}}}$ are equivalent to the first (when arranged by magnitude) $N_{\theta'}$ eigenvalues of $\bm{A}$ and are known as DMD eigenvalues \cite{proctor2016dynamic, schmid2010dynamic}.


As basis functions $\varphi_i$ in \eqref{eq:ROM_approx_t}, one takes the 
eigenvectors of $\bm{A}$ associated to the DMD eigenvalues. It can be shown \cite{tu2013dynamic} that these eigenvectors, called exact DMD modes, are the columns of the matrix $\mathbb{\bm{\varphi}} \in \mathbb R ^{N_c\times N_{\theta^{\prime}}}$:
\begin{align}
\mathbb{\bm{\varphi}} = \mathbb{\bm{S_2}} \mathbb{\bm{V}}_{N_{\theta^{\prime}}} \mathbb{\bm{\Sigma}}_{N_{\theta^{\prime}}}^{-1} \mathbb{\bm{W}}.
\label{eq:DMD4}
\end{align}
Another possibility (not explored in this paper) is to take the columns of 
$\mathbb{\bm{U}}_{N_{\theta^{\prime}}} \mathbb{\bm{W}}$, which are known 
as projected DMD modes \cite{schmid2010dynamic}.

Finally, the temporal coefficients $\alpha_i$ in eq. \eqref{eq:ROM_approx_t} are given by
\begin{align}
\mathbb{\alpha}_i(t) = e^{\lambda_i t}, \quad i = 1, \dots, N_{\theta'}.
\label{eq:DMD5}
\end{align}

\subsection{Hankel Dynamic Mode Decomposition }\label{sec:HDMD}

The Hankel DMD (\cite{arbabi2017ergodic,  curtis2023machine, vasconcelos2019dynamic, yang2021synchronized}) is a variation of the standard DMD algorithm based on the idea to combine the DMD algorithm with time delay-embedding \cite{arbabi2017ergodic, fujii2019data,jiang2015study}. In delay embedding, a given time-dependent datum, which in our case is the snapshot $\theta'_h(\bm{x},t)$, is augmented by a time history of previous data. 




Let $M$ be the embedding dimension and let $\bm{H}(\theta^{\prime}_h(\bm{x}, t^i)) \in \mathbb{R}^{N_c \times M}$, with $i = 1, \dots, N_t$ and $N_t<N_{tf}$, be the so-called Hankel matrix associated to the snapshot $\theta^{\prime}_h(\bm{x}, t^i)$:
\begin{align}
\bm{H}(\theta^{\prime}_h(\bm{x}, t^i)) = [\theta^{\prime}_h(\bm{x}, t^i), \theta^{\prime}_h(\bm{x}, t^{i-1}), \dots  \theta^{\prime}_h(\bm{x}, t^{i - (M-1)})].
\label{eq:H}
\end{align}
We build two matrices in $\mathbb{R}^{N_c \times \left( (N_t - M)M \right)}$:
\begin{align}
\mathbb{\bm{S}}_1^H= 
\begin{bmatrix}
    \bm{H}(\theta^{\prime}_h(\bm{x}, t^M)) & \bm{H}(\theta^{\prime}_h(\bm{x}, t^{M+1})) & \dots & \bm{H}(\theta^{\prime}_h(\bm{x}, t^{N_t -1})) 
\end{bmatrix},
    \notag \\
    \mathbb{\bm{S}}_2^H = 
\begin{bmatrix}
    \bm{H}(\theta^{\prime}_h(\bm{x}, t^{M+1}))& \bm{H}(\theta^{\prime}_h(\bm{x}, t^{M+2}))& \dots & \bm{H}(\theta^{\prime}_h(\bm{x}, t^{N_t}))
\end{bmatrix}.
\end{align}
Then, we follow the DMD algorithm described Sec.~\ref{sec:DMD} by simply replacing $\mathbb{\bm{S}}_1$ with $\mathbb{\bm{S}}_1^H$
and $\mathbb{\bm{S}}_2$ with $\mathbb{\bm{S}}_2^H$.

It should be noted that the embedding dimension $M$ is a crucial parameter affecting the accuracy of the HDMD algorithm. The optimal value of $M$ could depend on the problem at hand and, to the best of our knowledge, there is no general rule to choose it \cite{jiang2015study, golyandina2001analysis, muruganatham2013roller}. In this work, its value is determined through a trial and error procedure, i.e., we try several values of $M$ and choose the one that maximizes the accuracy of the associated HDMD-based ROM in the $L^2$ norm. See Sec.~\ref{sec:RTB} for more details. 
Of course, such a procedure is time-consuming and leaves room to further research for improvement. 




\subsection{Proper Orthogonal Decomposition with Interpolation}\label{subsec:PODI}

In the PODI method, which was introduced in \cite{bui2003proper}, the POD algorithm is used to extract the reduced basis functions from the set of full order solutions associated with given values of time and $\bm{\pi}$, whilst 
the coefficients $\alpha_i$ are approximated by an interpolation technique.  
In the following, we briefly recall the main steps required by PODI method. 

Following the notation introduced in  Sec.~\ref{sec:FOM} and \ref{sec:DMD}, let $\theta_h^{\prime} (\bm{x}; t^i, \bm{\pi}^j)
\in \mathbb{R}^{N_c}$, with $i = 1, \dots, N_{tf}$ and $j = 1, \dots, N_\pi$, be the full order solution computed at  time instant $t^i$ and for parameter value $\bm{\pi}^j$. 
We set $N_s = N_{tf} \cdot N_\pi$
and arrange these full order solutions as the columns of the snapshot matrix 
\begin{align}
\mathbb{\bm{S}} = [\theta_h^{\prime}(\bm{x}; t^1, \bm{\pi}^1), \theta_h^{\prime}(\bm{x}; t^2, \bm{\pi}^1), \dots, \theta_h^{\prime}(\bm{x}; t^{N_{tf}}, \bm{\pi}^{N_k})] \in \mathbb{R}^{N_c \times N_s}. \label{eq:S}
    \end{align}
    
The SVD of matrix $\mathbb{\bm{S}}$ gives us:
\begin{align}
\mathbb{\bm{S}} = \U \bm{\Sigma} \bm{V}^T,
\end{align}
where the explanation for $\U \in \mathbb R^{{N_c}\times {N_c}}$, $\bm{\Sigma} \in \mathbb R ^{{N_c}\times {N_s}}$, and 
$\bm{V} \in R^{{N_s}\times {N_s}}$ is provided in Sec.~\ref{sec:DMD}.
The POD space is constructed by retaining the first $N_{\theta'} \leq \min\{N_c, N_s\}$ columns of matrix $\U$. 
The value of $N_{\theta^{\prime}}$ is set as described in Sec.~\ref{sec:DMD} and, like in Sec.~\ref{sec:DMD}, we will denote with $\U_{N_{\theta^{\prime}}}$ 
the matrix that contains the first $N_{\theta'}$ columns of $\U$.





Now that we have the basis functions, we can use eq. \eqref{eq:ROM_1} to approximate the snapshots: 
\begin{align}
\theta_h^{\prime} (\bm{x}; t^i, \bm{\pi}^j) \approx
\theta_r^{\prime}(\bm{x}; t^i, \bm{\pi}^j) = \sum_{L=1}^{N_{\theta^{\prime}}}\alpha_L (t^i, \bm{\pi}^j) \varphi_L (\bm{x}),  \label{eq:snap_approx}
\end{align}
for $i = 1, \dots, N_{tf}$ and $j = 1, \dots, N_\pi$, where the coefficients $\alpha_L (t^i, \bm{\pi}^j)$ are the 
entries of the matrix $\bm{C} = \U_{N_{\theta^{\prime}}}^T \bm{S} \in \mathbb{R}^{N_{\theta^{\prime}} \times N_s}$, i.e., $\alpha_L (t^i, \bm{\pi}^j) = \bm{C}_{ij}$.
Given these coefficients at $t^i$
and $\bm{\pi}^j$, we construct an interpolant with 
Radial Basis Functions \cite{buhmann2003radial} as follows:
\begin{align}
    \alpha_L(t^{i}, \bm{\pi}^{j}) = \sum_{m=1}^{N_{tf}} \sum_{n=1}^{N_\pi} w_{L_{m,n}} \zeta_{L_{m,n}}(\| (t^{i}, \bm{\pi}^{j}) - (t^m, \bm{\pi}^n) \|). \label{Al}
\end{align}
Here, $\zeta_{L_{m,n}}$ are the Radial Basis Functions (choosen to be Gaussian functions) centered in $(t^m, \bm{\pi}^n)$ and $w_{L_{m,n}}$ are unknown weights. We find the weights by solving the linear system associated to \eqref{Al}:
\begin{align}
\bm{Z}_L \bm{w}_L = \bm{\alpha}_L. 
\label{sys}
\end{align}
We remark that it is done only once during the offline phase.

Let us now consider a solution that does not belong to the snapshot matrix, i.e., we want to 
compute $\theta^{\prime}(\bm{x}; t^{new}, \bm{\pi}^{new})$ for new time $t^{new}$ and new parameter value $\bm{\pi}^{new}$. We obtain
\begin{align}
\theta_r^{\prime}(\bm{x}; t^{new}, \bm{\pi}^{new}) = \sum_{L=1}^{N_{\theta'}} \alpha_L(t^{new}, \bm{\pi}^{new})\varphi_L (\bm{x}), 
\end{align} 
where coefficients $\alpha_L(t^{new}, \bm{\pi}^{new})$ are computed by:
\begin{align}
    \alpha_L(t^{new}, \bm{\pi}^{new}) = \sum_{i=1}^{N_{tf}} \sum_{j=1}^{N_\pi} w_{L_{m,n}} \zeta_{L_{m,n}}(\| (t^{new}, \bm{\pi}^{new}) - (t^i, \bm{\pi}^j) \|). \label{Al2}
\end{align}
Notice that while \eqref{Al} is used to find the weights $w_{L_{m,n}}$, 
\eqref{Al2} is used to find $\alpha_L$. 

\section{Numerical results}\label{sec:res}
\graphicspath{{./img/}}

To validate our ROM approach, we consider two standard benchmarks for atmospheric flows: the rising thermal bubble \cite{ahmad2007euler,ahmad2018high,feng2021hybrid,marras2015stabilized} and the density current \cite{ahmad2007euler,carpenter1990application,giraldo2008study,marras2013variational,marras2015stabilized,straka1993numerical}. Both test cases involve a perturbation of a neutrally stratified atmosphere with
uniform background potential temperature.
It is worth to note that there exist several variations of these benchmarks, featuring different geometries and/or initial conditions. We use the
setting from \cite{ahmad2007euler}
for the rising thermal bubble and the setting from \cite{carpenter1990application, straka1993numerical} for the density current.

We compare the ROM techniques presented in Sec.~\ref{sec:ROM} in terms of the reconstruction of the time evolution of the potential temperature perturbation for the rising thermal bubble in 
Sec. \ref{sec:RTB} and for the density current in Sec. \ref{sec:DC}. Moreover, in Sec. \ref{sec:DC} we present a parametric study with  PODI, where the varying parameter is the
amplitude of the initial temperature perturbation 
in the density current benchmark. 

The reader interested in the validation of the FOM model is referred to \cite{girfoglio2023validationAIP}.

\subsection{Rising thermal bubble}\label{sec:RTB}

The computational domain for this benchmark is $\Omega = [0,5000] \times [0, 10000]$ m$^2$ in the $xz$-plane. In this domain, 
a neutrally stratified atmosphere with uniform background potential temperature $\theta_0$=300 K is perturbed by a circular bubble of warmer air.
The initial temperature field is
\begin{equation}
\theta^0 = 300 + 2\left[ 1 - \frac{r}{r_0} \right] ~ \textrm{if $r\leq r_0=2000~\mathrm{m}$}, \quad\theta^0 = 300
~ \textrm{otherwise},
\label{warmEqn1}
\end{equation}
where $(x_c,z_c) = (5000,2000)~\mathrm{m}$ and $r = \sqrt[]{(x-x_{c})^{2} + (z-z_{c})^{2}}$ are the center and the radius of the circular perturbation, respectively \cite{ahmad2007euler,ahmad2018high}.
The initial density is given by 
\begin{align}
\rho^0 = \frac{p_g}{R \theta_0} \left(\frac{p}{p_g}\right)^{c_{v}/c_p} \quad \text{with} \quad p = p_g \left( 1 - \frac{g z}{c_p \theta^0} \right)^{c_p/R}, \label{eq:rho_wb}
\end{align}
where 
$c_p = R + c_v$, with $c_v = 715.5$ J/(Kg K) and $R = 287$ J/(Kg K). 
The initial velocity field is zero everywhere. 
The initial specific enthalpy is defined as:
\begin{align}
h^{0} = c_p \theta^0 \left( \frac{p}{p_g} \right)^{\frac{R}{c_{p}}}.
\label{eq:e0}
\end{align}
We impose impenetrable, free-slip boundary conditions on all the boundaries and we set $t_f = 1020$ s. In time interval $(0, 1020]$ s, the warm bubble rises due to buoyancy and evolves into a mushroom-like shape as a result of shear stress.

We generate a uniform structured mesh with mesh size $h = \Delta x = \Delta z = 62.5$ m and set the time step set to $\Delta t = 0.1$ s. Following \cite{ahmad2007euler,girfoglio2023validationAIP}, 
we set $\mu_a = 15$ and $Pr = 1$ in \eqref{eq:mom_LES}-\eqref{eq:ent_LES}.
Both of these are ah-hoc values provided in \cite{ahmad2007euler} to stabilize the numerical simulations. More sophisticated LES models can be found in, e.g., \cite{clinco2023filter, marras2015stabilized}.

We collect an original database consisting of 204 snapshots, i.e., the computed $\theta^\prime$ every 5 seconds. These snapshots are divided into two different sets. A first set, called training set, is used to generate the reduced basis. All the snapshots belonging to the training set are stored in the matrix $\mathbb{\bm{S}}$ for PODI, while all but one are stored in 
$\mathbb{\bm{S}}_1$ for DMD, and $\mathbb{\bm{S}}^H_1$ for HDMD
as explained in Sec.~\ref{sec:DMD} and
\ref{sec:HDMD}.
The second
set, called validation set, is the complement of the training set in the original database and it is used to assess the accuracy of the ROM solution.
The partitioning into these sets can be done in two ways: randomly or by preserving the temporal order. 

Out of the 204 computed $\theta'$ in the original database, we take 184 (i.e., 90\% of the database) to form the training set. In the case of DMD and Hankel DMD, these 184 solutions are the first 184 in the database (associated to the time interval $(0, 920]$ s). In the case of PODI instead, these 184 solutions are selected randomly over the entire time interval $[0, 1020]$ s. For all three methods, 
the remaining 20 solutions form the validation set. This difference in the training and validation sets reflects the different nature of PODI and DMD/HDMD algorithms: the former relies on the reconstruction of the solution, the latter predicts the future behavior of a system. 

\begin{figure}
    \centering \includegraphics[width=80mm,scale=0.5]{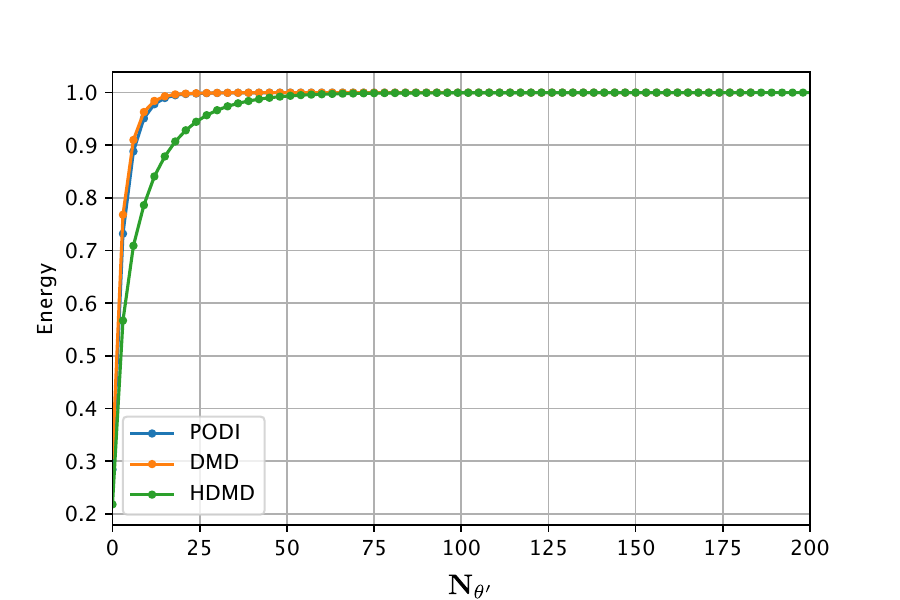}
        \caption{Rising thermal bubble:  cumulative energy  $E$ in \eqref{k_equation} for the three ROM approaches under consideration.}
        \label{fig:modes_energy_theta_RTB}
\end{figure}

We start with the plot of the cumulative eigenvalues energy $E$ \eqref{k_equation} for PODI, DMD and HDMD in \fig{fig:modes_energy_theta_RTB}. We see that the curves for PODI and DMD are very close, while the curve for HDMD is farther apart. In particular,
HDMD requires a larger number of singular values to reach a given energy level. To clarify the extent of this difference,  Tab.~\ref{tab:RTB_Modes} reports the number of modes needed to attain $\delta = 0.7, 0.9, 0.99$ in \eqref{k_equation}.
We observe that DMD and PODI require the same number of modes for a given $\delta$,  while HDMD needs
about or more than twice as many modes. 

\begin{table}[ht!]
\centering
\begin{tabular}{|c|c|c|c|}
\hline  
   & DMD  & HDMD &   PODI \\ \hline
$\delta = 70 \%$ & 4  &  7  & 4  \\ \hline
$\delta = 90 \%$ & 8  &  19 & 8 \\ \hline
$\delta = 99 \%$ & 17 &  46 & 17 \\\hline 
\end{tabular}
  \caption{Rising thermal bubble: number of modes required to retain different energy thresholds, $\delta = 0.7, 0.9, 0.99$, for the three ROM methods under consideration.
  }
 \label{tab:RTB_Modes}
\end{table}

\begin{figure}
\vspace{1cm}
  \begin{overpic}[width=.49\textwidth, grid=false]{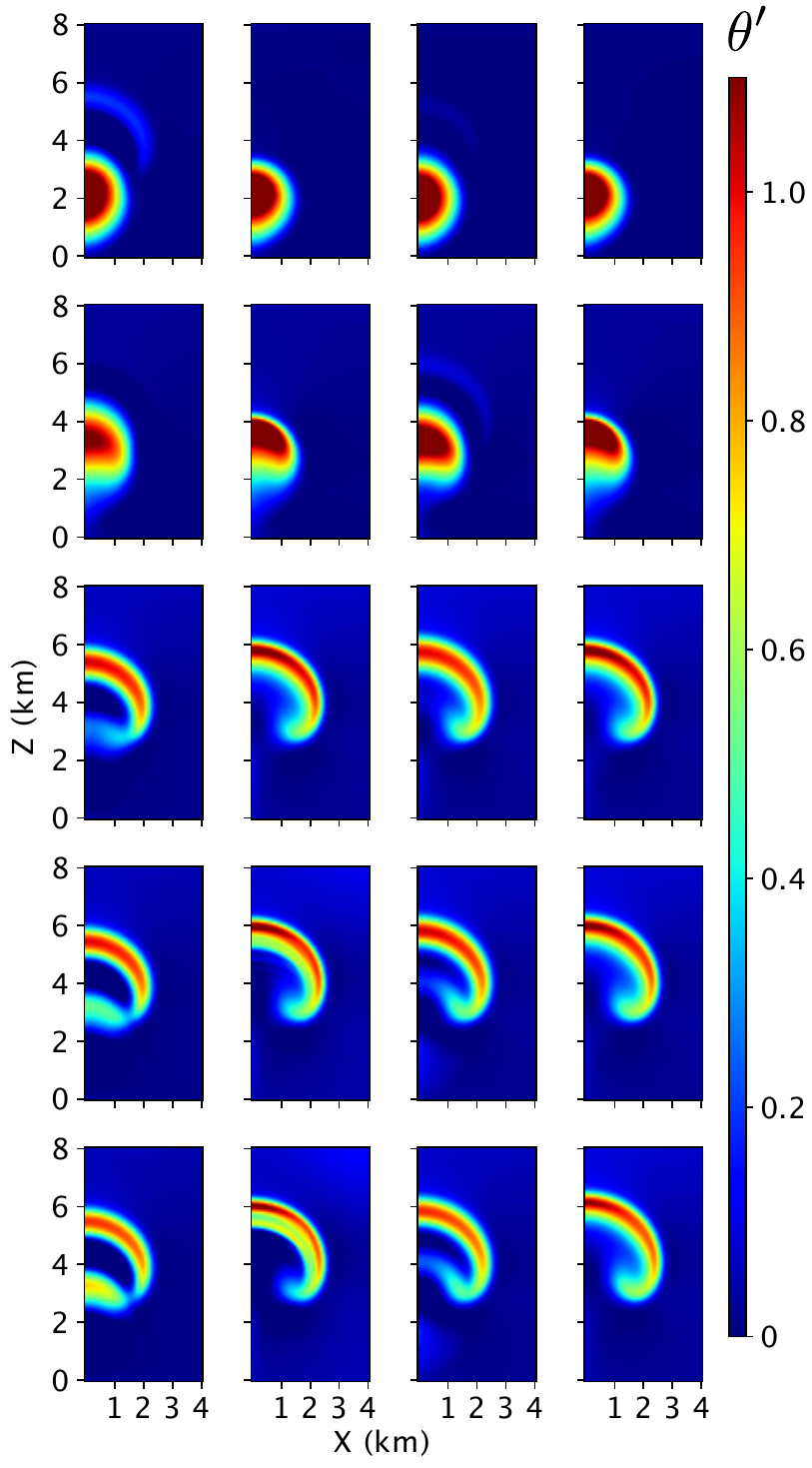}  
        \put(22,104){$\delta = 0.7$}
         \put(6.5,99.5){DMD}
         \put(16.8,99.5){HDMD}
         \put(28.8,99.5){PODI}
         \put(40.2,99.5){FOM}
         \put(6.7,96.5){\tiny{\textcolor{white}{$t$=255 s}}}
         \put(6.7,77.5){\tiny{\textcolor{white}{$t$=505 s}}}
         \put(6.7,58.5){\tiny{\textcolor{white}{$t$=930 s}}}
         \put(6.7,39.3){\tiny{\textcolor{white}{$t$=980 s}}}
         \put(6,20.2){\tiny{\textcolor{white}{$t$=1020 s}}}

          \put(18,96.5){\tiny{\textcolor{white}{$t$=255 s}}}
         \put(18,77.5){\tiny{\textcolor{white}{$t$=505 s}}}
         \put(18,58.5){\tiny{\textcolor{white}{$t$=930 s}}}
         \put(18,39.3){\tiny{\textcolor{white}{$t$=980 s}}}
         \put(17.3,20.2){\tiny{\textcolor{white}{$t$=1020 s}}}

          \put(29.3,96.5){\tiny{\textcolor{white}{$t$=255 s}}}
         \put(29.3,77.5){\tiny{\textcolor{white}{$t$=505 s}}}
         \put(29.3,58.5){\tiny{\textcolor{white}{$t$=930 s}}}
         \put(29.3,39.3){\tiny{\textcolor{white}{$t$=980 s}}}
         \put(28.7,20.2){\tiny{\textcolor{white}{$t$=1020 s}}}

          \put(40.5,96.5){\tiny{\textcolor{white}{$t$=255 s}}}
         \put(40.5,77.5){\tiny{\textcolor{white}{$t$=505 s}}}
         \put(40.5,58.5){\tiny{\textcolor{white}{$t$=930 s}}}
         \put(40.5,39.3){\tiny{\textcolor{white}{$t=$980 s}}}
         \put(40.2,20.2){\tiny{\textcolor{white}{$t$=1020 s}}}
      \end{overpic}
    \begin{overpic}[width=.49\textwidth, grid=false]{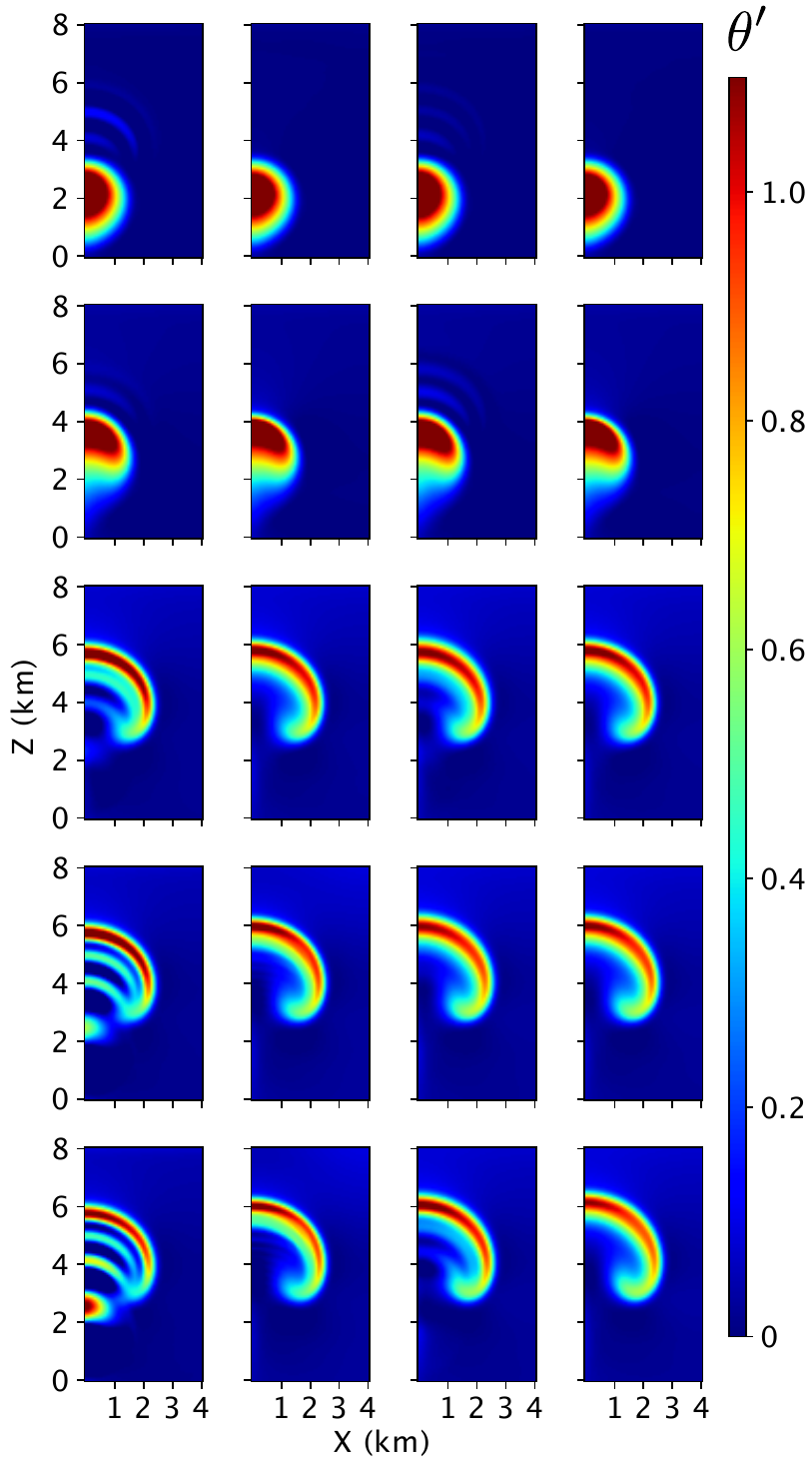}  
        \put(22,104){$\delta = 0.9$}
         \put(6.5,99.5){DMD}
         \put(16.8,99.5){HDMD}
         \put(28.8,99.5){PODI}
         \put(40.2,99.5){FOM}
         \put(6.7,96.5){\tiny{\textcolor{white}{$t$=255 s}}}
         \put(6.7,77.5){\tiny{\textcolor{white}{$t$=505 s}}}
         \put(6.7,58.5){\tiny{\textcolor{white}{$t$=930 s}}}
         \put(6.7,39.3){\tiny{\textcolor{white}{$t$=980 s}}}
         \put(6,20.2){\tiny{\textcolor{white}{$t$=1020 s}}}

          \put(18,96.5){\tiny{\textcolor{white}{$t$=255 s}}}
         \put(18,77.5){\tiny{\textcolor{white}{$t$=505 s}}}
         \put(18,58.5){\tiny{\textcolor{white}{$t$=930 s}}}
         \put(18,39.3){\tiny{\textcolor{white}{$t$=980 s}}}
         \put(17.3,20.2){\tiny{\textcolor{white}{$t$=1020 s}}}

          \put(29.3,96.5){\tiny{\textcolor{white}{$t$=255 s}}}
         \put(29.3,77.5){\tiny{\textcolor{white}{$t$=505 s}}}
         \put(29.3,58.5){\tiny{\textcolor{white}{$t$=930 s}}}
         \put(29.3,39.3){\tiny{\textcolor{white}{$t$=980 s}}}
         \put(28.7,20.2){\tiny{\textcolor{white}{$t$=1020 s}}}

          \put(40.5,96.5){\tiny{\textcolor{white}{$t$=255 s}}}
         \put(40.5,77.5){\tiny{\textcolor{white}{$t$=505 s}}}
         \put(40.5,58.5){\tiny{\textcolor{white}{$t$=930 s}}}
         \put(40.5,39.3){\tiny{\textcolor{white}{$t=$980 s}}}
         \put(40.2,20.2){\tiny{\textcolor{white}{$t$=1020 s}}}
      \end{overpic}
   \caption{Rising thermal bubble: comparison of the evolution of $\theta'$
   given by the ROMs (first 3 columns in each panel) and the FOM (last column in each panel) for $\delta = 0.7$ (left panel) and $\delta = 0.9$ (right panel).} 
    \label{fig:Vis_POD_DMD_HDMD_RTB_70_90}
\end{figure}

\begin{figure}
    \vspace{1cm}
    \centering
    \begin{overpic}[width=0.48\textwidth, grid=false]{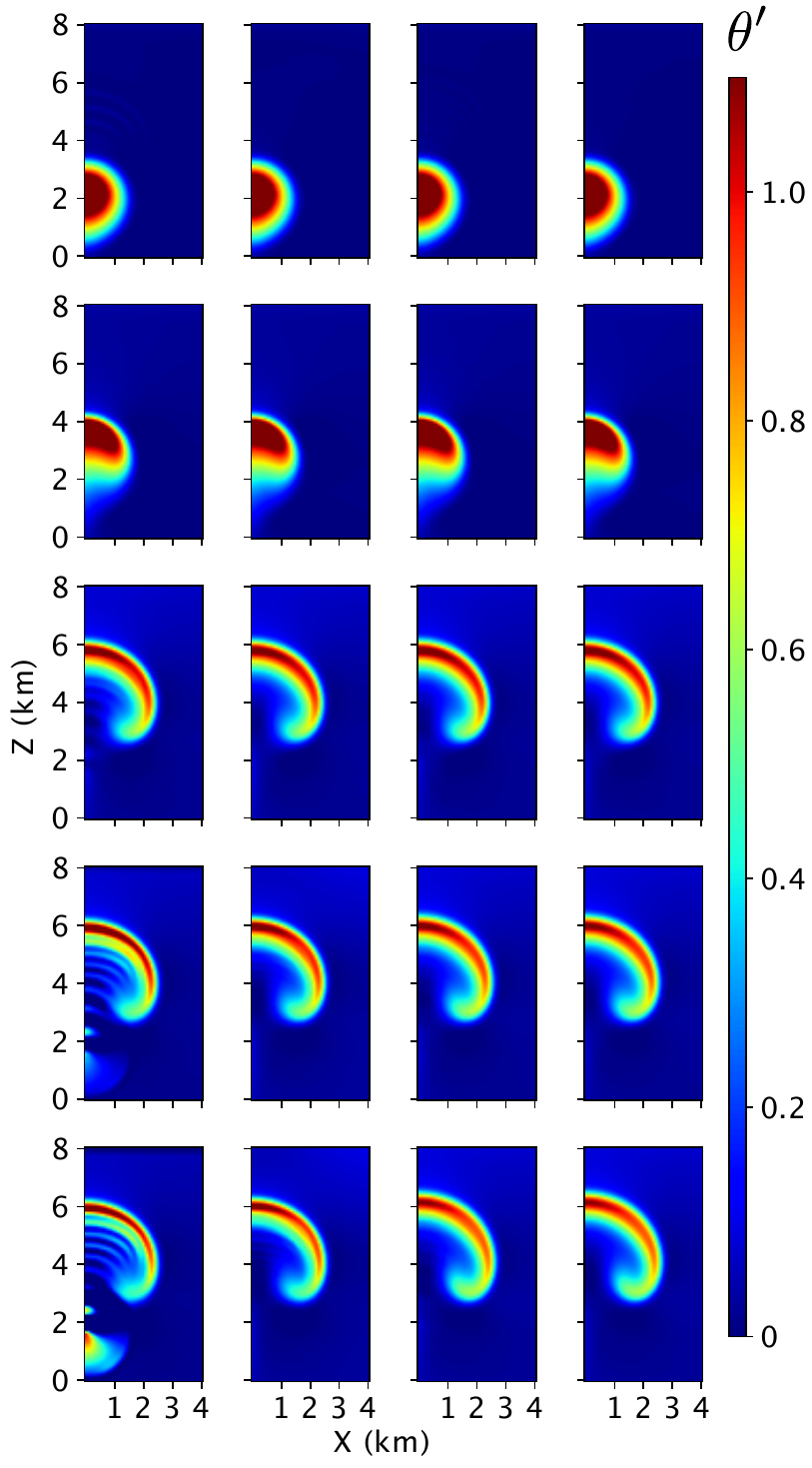}  
        \put(22,104){$\delta = 0.99$}
         \put(6.5,99.5){DMD}
         \put(16.8,99.5){HDMD}
         \put(28.8,99.5){PODI}
         \put(40.2,99.5){FOM}
         \put(6.7,96.5){\tiny{\textcolor{white}{$t$=255 s}}}
         \put(6.7,77.5){\tiny{\textcolor{white}{$t$=505 s}}}
         \put(6.7,58.5){\tiny{\textcolor{white}{$t$=930 s}}}
         \put(6.7,39.3){\tiny{\textcolor{white}{$t$=980 s}}}
         \put(6,20.2){\tiny{\textcolor{white}{$t$=1020 s}}}
          \put(18,96.5){\tiny{\textcolor{white}{$t$=255 s}}}
         \put(18,77.5){\tiny{\textcolor{white}{$t$=505 s}}}
         \put(18,58.5){\tiny{\textcolor{white}{$t$=930 s}}}
         \put(18,39.3){\tiny{\textcolor{white}{$t$=980 s}}}
         \put(17.3,20.2){\tiny{\textcolor{white}{$t$=1020 s}}}

          \put(29.3,96.5){\tiny{\textcolor{white}{$t$=255 s}}}
         \put(29.3,77.5){\tiny{\textcolor{white}{$t$=505 s}}}
         \put(29.3,58.5){\tiny{\textcolor{white}{$t$=930 s}}}
         \put(29.3,39.3){\tiny{\textcolor{white}{$t$=980 s}}}
         \put(28.7,20.2){\tiny{\textcolor{white}{$t$=1020 s}}}

          \put(40.5,96.5){\tiny{\textcolor{white}{$t$=255 s}}}
         \put(40.5,77.5){\tiny{\textcolor{white}{$t$=505 s}}}
         \put(40.5,58.5){\tiny{\textcolor{white}{$t$=930 s}}}
         \put(40.5,39.3){\tiny{\textcolor{white}{$t=$980 s}}}
         \put(40.2,20.2){\tiny{\textcolor{white}{$t$=1020 s}}}
      \end{overpic}
      \captionsetup{list=no} 
  \caption{Rising thermal bubble: comparison of the evolution of $\theta'$
   given by the ROMs (first 3 columns) and the FOM (last column) for $\delta = 0.99$.}
   \label{fig:Vis_POD_DMD_HDMD_RTB_99}
\end{figure}

Next, Fig.~\ref{fig:Vis_POD_DMD_HDMD_RTB_70_90} and \ref{fig:Vis_POD_DMD_HDMD_RTB_99} show a qualitative comparison of the ROM solutions for $\delta= 0.7, 0.9, 0.99$ with the FOM solution. Among the five time instants chosen for the visualization, two of them, namely $t = 255$ s and $t = 505$ s, correspond to solutions belonging to the training set. These two times allow us to asses the ability of each ROM technique to identify the system dynamics. The remaining three times, i.e., $t = 930,980, 1020$ s, are not associated with the training set and thus are used to check the accuracy of the ROM in predicting (for DMD and HDMD) or interpolating (for PODI) the system dynamics. 
Let us discuss the results in Fig.~\ref{fig:Vis_POD_DMD_HDMD_RTB_70_90} and \ref{fig:Vis_POD_DMD_HDMD_RTB_99} starting from the system identification. For $\delta = 0.7, 0.9$, HDMD provides a better reconstruction of $\theta'$ than PODI and DMD. See the first two rows in 
Fig.~\ref{fig:Vis_POD_DMD_HDMD_RTB_70_90}. However, when $\delta$ is increased to 0.99 we observe that all three ROMs reconstruct $\theta'$ well. Indeed, no significant difference can be observed in the panels of row one and two of 
Fig.~\ref{fig:Vis_POD_DMD_HDMD_RTB_99}.
Now, let us take a look at the solutions corresponding to the times not associated with the training set. 
From the bottom three rows in Fig.~\ref{fig:Vis_POD_DMD_HDMD_RTB_70_90} and \ref{fig:Vis_POD_DMD_HDMD_RTB_99}, we see that for none of the values of $\delta$ DMD can correctly predict the evolution of $\theta'$, while HDMD and PODI reconstruct well
the solution when $\delta$ is set to $0.99$.

\begin{figure}
\centering
\begin{overpic}[width=0.48\textwidth, grid=false]{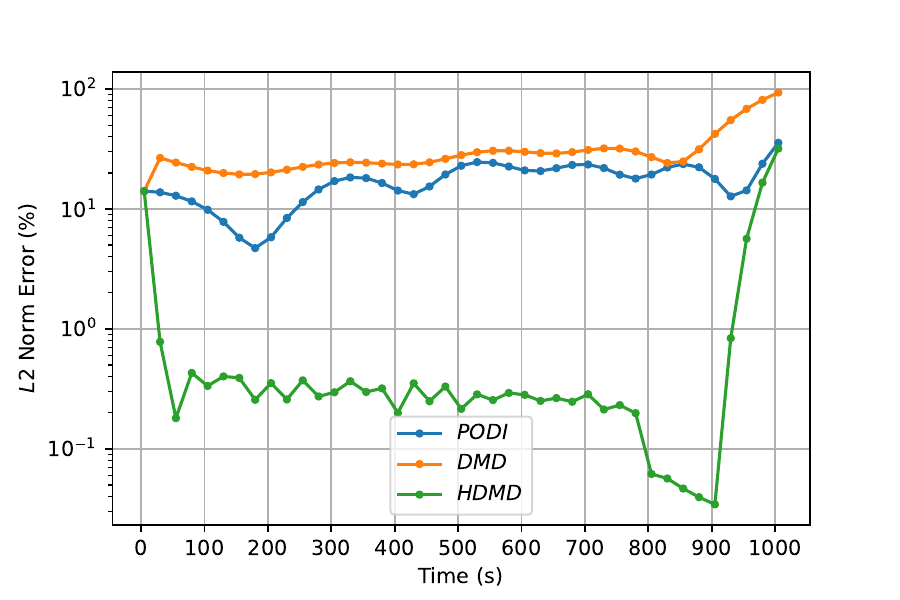}  
        \put(43,62){$\delta = 0.7$}
\end{overpic}
\begin{overpic}[width=0.48\textwidth, grid=false]{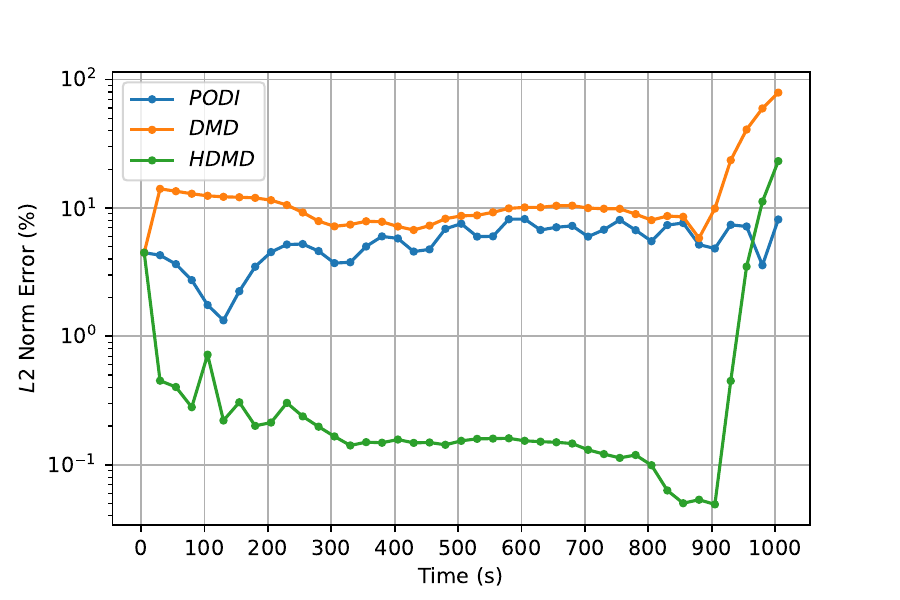}  
        \put(43,62){$\delta = 0.9$}
\end{overpic}\\
\begin{overpic}[width=0.48\textwidth, grid=false]{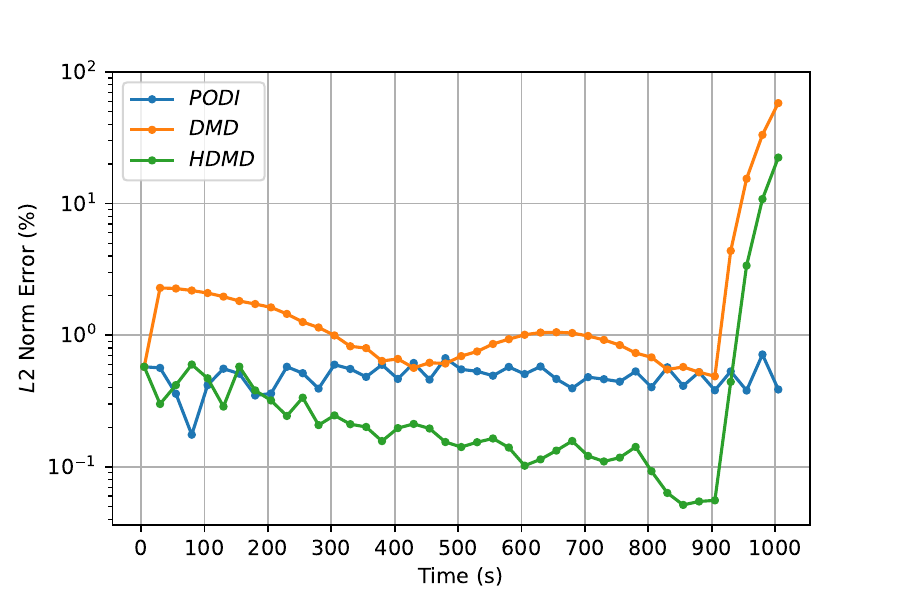} 
    \put(43,62){$\delta = 0.99$}
\end{overpic} 
\caption{Rising thermal bubble: time evolution of the $L^2$ error \eqref{eq:l2Error} for DMD, HDMD, and PODI for $\delta= 0.7$ (top left panel), $\delta= 0.9$ (top right panel) and $\delta= 0.99$ (bottom panel).}
\label{fig:Error_in_time_POD_DMD_HDMD_RTB}
\end{figure}

To quantitate the agreement between the ROM solutions and the FOM solutions in Fig.~\ref{fig:Vis_POD_DMD_HDMD_RTB_70_90} and \ref{fig:Vis_POD_DMD_HDMD_RTB_99}, we report 
in \fig{fig:Error_in_time_POD_DMD_HDMD_RTB} 
the $L^2$ error 
\begin{equation}
E_{\theta^\prime}(t) = 100 \cdot \dfrac{||\theta^\prime_h(t) - \theta^\prime_r(t)||_{L^2(\Omega)}}{||{\theta^\prime_h}(t)||_{L^2(\Omega)}},
\label{eq:l2Error}
\end{equation}
for the three values of $\delta$ under consideration. We see that the $L^2$ errors for DMD and HDMD increase sharply around and after $t = 920$ s for all values of $\delta$. 
Since the training set for DMD and HDMD includes only solutions before $t = 920$ s, an increase in the errors past that time is to be expected. However, the magnitude of the errors at the end of the time interval, which in the best case (HDMD with $\delta = 0.99$) is around 20\%, indicates that both DMD and HDMD do a poor job in predicting the evolution of the warm bubble. This is somewhat surprising since visually the HDMD solution for $\delta = 0.99$ compares well with the FOM solution for all the times shown in \fig{fig:Vis_POD_DMD_HDMD_RTB_99}, including the times corresponding to the validation set. 
We suspect that the larger error for HDMD is due to a difference in rising speed for the bubble, and not so much in the bubble shape.
Since the snapshots in the training dataset for PODI were chosen randomly, and thus include FOM solutions past $t = 920$ s, the $L^2$ errors for PODI are comparable to (for $\delta = 0.7$) or smaller (for $\delta = 0.9, 0.99$) than the errors for DMD and HDMD
when $t > 920$ s. In particular, we notice that then $\delta = 0.99$ the $L^2$ errors for PODI remain less than 1\%  for the entire duration of the time interval. In time interval $(0, 920]$ s, which corresponds to system identification, HDMD reconstructs the solution more accurately than DMD and PODI. This is especially true for $\delta = 0.7, 0.9$.



We point out that for HDMD we set $M=25$. 
We tried $M=5, 10, 25, 50,75$ and
found that this value 
provided a reasonable trad-off between accuracy measured with error \eqref{eq:l2Error} and computational cost, mentioned below.

We ran all the simulations simulations on a 11th Gen Intel$(R)$ Core(TM) i7-11700 $@$ 2.50GHz system with 32GB RAM.
The time required by the FOM to complete the simulation is 65 s.  Tab.~\ref{tab:cost} reports the computational time needed to construct the reduced basis offline and to perform a simulation online for each of the ROM we consider. 
The computational times for DMD and PODI are comparable, which for the online phase is explained by the fact that the DMD and PODI reduced basis have the same size for the three values of $\delta$ we consider (see Tab.~\ref{fig:modes_energy_theta_RTB}).
The speed up, i.e., the ratio between the time for a FOM simulation and the online time for the ROM, is of the order of 3000 for these two methods. HDMD is computationally more expensive than DMD and PODI, which is the price one has to pay for the increased accuracy in system identification (see Fig.~\ref{fig:Error_in_time_POD_DMD_HDMD_RTB}). The speed up for HDMD is only about 20.



\begin{table}
\centering
\begin{tabular}{|c|c|c|c|c|c|c|}
\hline 
  & \multicolumn{2}{c|}{DMD}  & \multicolumn{2}{c|}{HDMD}  &   \multicolumn{2}{c|}{PODI} \\
\hline
$\delta $ & Basis & Online & Basis & Online & Basis & Online  \\ \hline
 $70 \%$ &  0.0833 s & 0.0182 s & 2.487 s & 2.73 s  & 0.074 s & 0.012 s   \\ 
 \hline
 $90 \%$ &  0.0835 s & 0.0193 s & 2.546 s & 2.822 s & 0.092 s & 0.015 s  \\ \hline
 $99 \%$ & 0.085 s  & 0.02 s   & 2.865 s & 2.95 s  & 0.1 s   & 0.018 s  \\\hline
\end{tabular}
 \caption{Rising thermal bubble: 
 computational time needed to construct the reduced basis offline (Basis) and to perform a simulation online (Online) for DMD, HDMD, and PODI when $\delta$ is set to $0.7, 0.9, 0.99$.}
 \label{tab:cost}
\end{table}
\subsection{Density current}\label{sec:DC}

\begin{figure}
    \centering \includegraphics[width=0.6\textwidth]{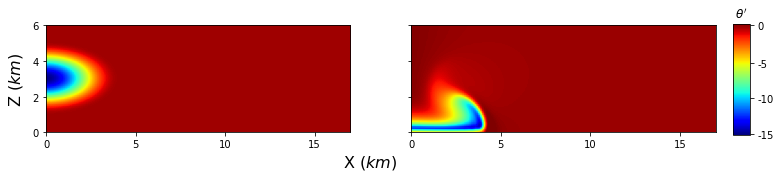}
        \caption{Density current: initial condition (left) and computed solution at $t=280$ s (right). }    
        \label{fig:DC_InitialCon}
\end{figure}

The computational domain for this benchmark is $\Omega = 25600 \times 6400$ m$^2$ in $xz$-plane. Like for the previous benchmark, we start from a neutrally stratified atmosphere with uniform background potential temperature $\theta_0$=300 K by introducing a perturbation. However, unlike the previous benchmark, the perturbation 
is represented by a circular bubble of colder air. The initial temperature field is
\begin{align}
    \theta_0 = 300 - \theta_s [1+ \cos(\pi r)], \quad \mathrm{if} \ r\le1, \quad \mathrm{otherwise} \ \theta_0 = 300, 
    \label{eq:initila_pot_temp_DC}
\end{align}
where $r = \sqrt[]{\left(\frac{x-x_{c}}{x_r}\right)^{2} + \left(\frac{z-z_{c}}{z_r}\right)^{2}}$, with $(x_r,z_r)=(4000, 2000)~{\rm m}$ and $(x_c,z_c) = (0,3000)~\mathrm{m}$. In \eqref{eq:initila_pot_temp_DC}, $\theta_s$ is the semi-amplitude of the initial temperature perturbation. In \cite{ahmad2007euler,carpenter1990application,giraldo2008study,marras2013variational,marras2015stabilized,straka1993numerical}, $\theta_s$ is set to 7.5. We will start from this value as well, but later will let $\theta_s$ vary in a given interval. The initial density is given by \eqref{eq:rho_wb}, while the initial specific enthalpy is \eqref{eq:e0}. The initial velocity field is zero everywhere.
We impose impenetrable, free-slip boundary conditions on all the walls. In
the time interval of interest, which is $(0, 900]$ s, the cold air descends due to negative buoyancy and when it reaches the ground, it rolls up and forms a front. As this front propagates, a multi-rotor structure develops.
One important difference with respect to the previous benchmark is that in this test the flow structures have a predominantly vertical motion till about $t = 280$ s (fall of the cold air) and then the motion becomes predominantly horizontal (front propagation). See \fig{fig:DC_InitialCon} for the initial solution and computed solution at $t=280$ s.

We generate a uniform structured mesh with mesh size $h = \Delta x = \Delta z = 100$ m and set the time step set to $\Delta t = 0.1$ s. Following \cite{ahmad2007euler,straka1993numerical}, 
we set $\mu_a = 75$ and $Pr = 1$ in \eqref{eq:mom_LES}-\eqref{eq:ent_LES}. Like in the case of the previous benchmark, these are ah-hoc values used to stabilize the numerical simulations.

We will compare the ROM techniques under consideration in terms of the reconstruction of the time evolution of $\theta'$ in Sec.~\ref{sec:Time_rec_DC}. 
In addition, in Sec.~\ref{Parametric_rec_DC}
we perform a parametric study for $\theta_s$
with PODI. 

\subsubsection{Time reconstruction}
\label{sec:Time_rec_DC}

For the results in this section, we set $\theta_s = 7.5$.  To generate the reduced basis, we collect a database consisting of 225 snapshots, i.e., the computed $\theta^\prime$ every 4 seconds. Similarly to the procedure reported in Sec.~\ref{sec:RTB}, we divide the database into two sets: a training set containing 90$\%$ of the snapshots (i.e., 202 snapshots) and a validation set containing the remaining 10\% (i.e., 23 snapshots). The partition of the database is performed sequentially for DMD and HDMD and corresponds to the first 202 snapshots. In the case of PODI, the 202 snapshots in the training set are selected randomly from the entire time interval. For HDMD, we set $M = 50$, which is larger that the value we used for the warm bubble because the density current flow is more complex.

\begin{figure}
    \centering \includegraphics[width=80mm,scale=0.5]{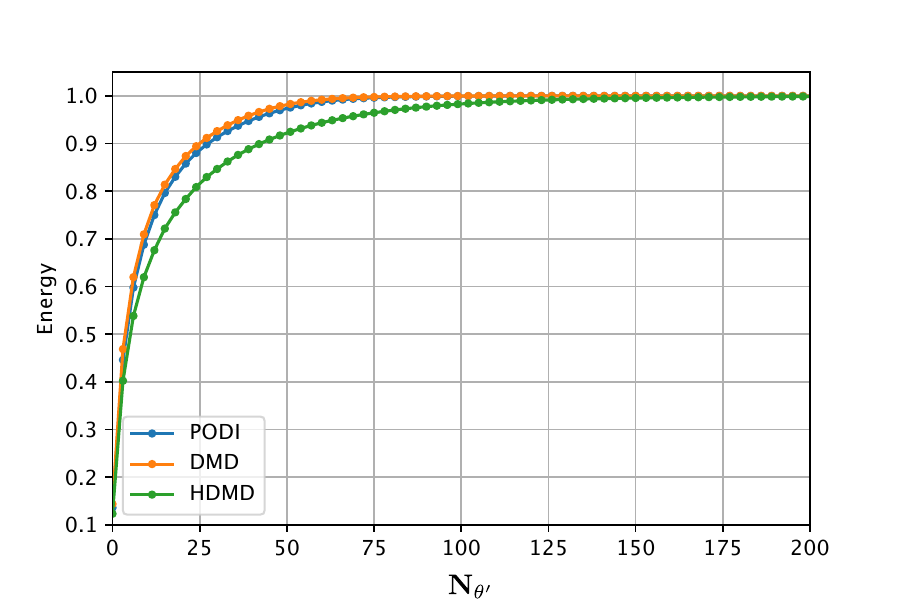}
        \caption{Density current: cumulative energy  $E$ in \eqref{k_equation} for the three ROM approaches under consideration.}
        \label{fig:modes_energy_theta_DC}
\end{figure}

The cumulative eigenvalue energy \eqref{k_equation} as the number of eigenvalues
is increased is shown in \fig{fig:modes_energy_theta_DC} for all the ROM approaches. We observe a less steep increase than in \fig{fig:modes_energy_theta_RTB}, indicating that more modes are required to capture the flow dynamics in the density current test than in the thermal rising bubble test. 
Tab.~\ref{tab:DC_1st_Modes} displays the number of modes required to capture different energy thresholds ($\delta = 0.7, 0.9, 0.99$) for PODI, DMD, and HDMD. PODI and DMD require the same number of modes for $\delta = 0.7$ and similar numbers for $\delta = 0.9, 0.99$. The number of modes needed by HDMD is larger for every $\delta$. While the observations for Tab.~\ref{tab:DC_1st_Modes} are similar to the observation for Tab.~\ref{tab:RTB_Modes}, the numbers in Tab.~\ref{tab:DC_1st_Modes} are larger. Again, 
this is due to the fact that the density current produces a more complex flow.

\begin{table}[htb!]
\centering
\begin{tabular}{|c|c|c|c|}
\hline  
   & DMD  & HDMD &   PODI \\ \hline
 $\delta = 70 \%$ & 11  &  15  & 11  \\ \hline
 $\delta = 90 \%$ & 28  &  44 & 29 \\ \hline
$\delta = 99 \%$ & 63 &  120 & 65 \\\hline
\end{tabular}
  \caption{Density current: number of modes required to retain different energy tresholds, $\delta= 0.7, 0.9, 0.99$, for three ROM methods under consideration.
  }
 \label{tab:DC_1st_Modes}
\end{table}

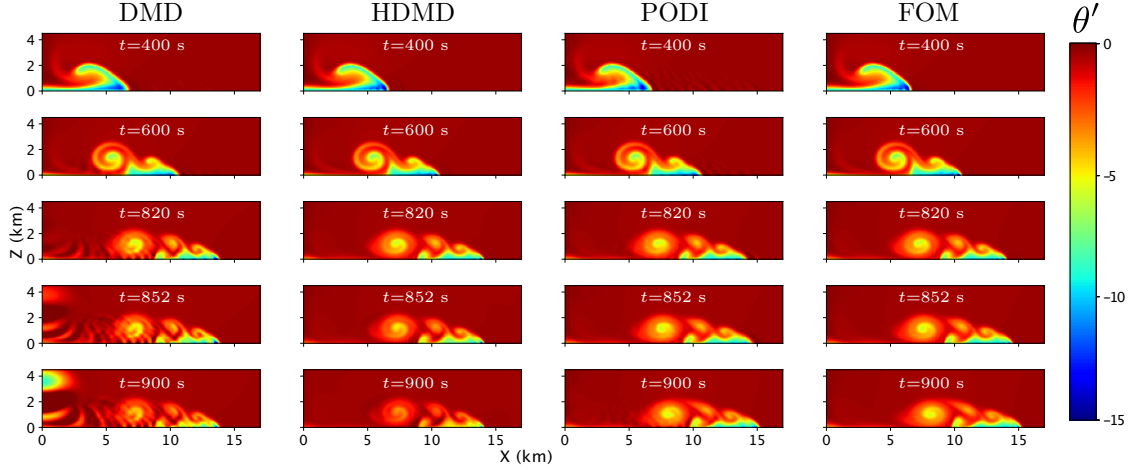
\begin{figure}
    \centering
    \begin{overpic}[width=1.0\textwidth, grid=false]{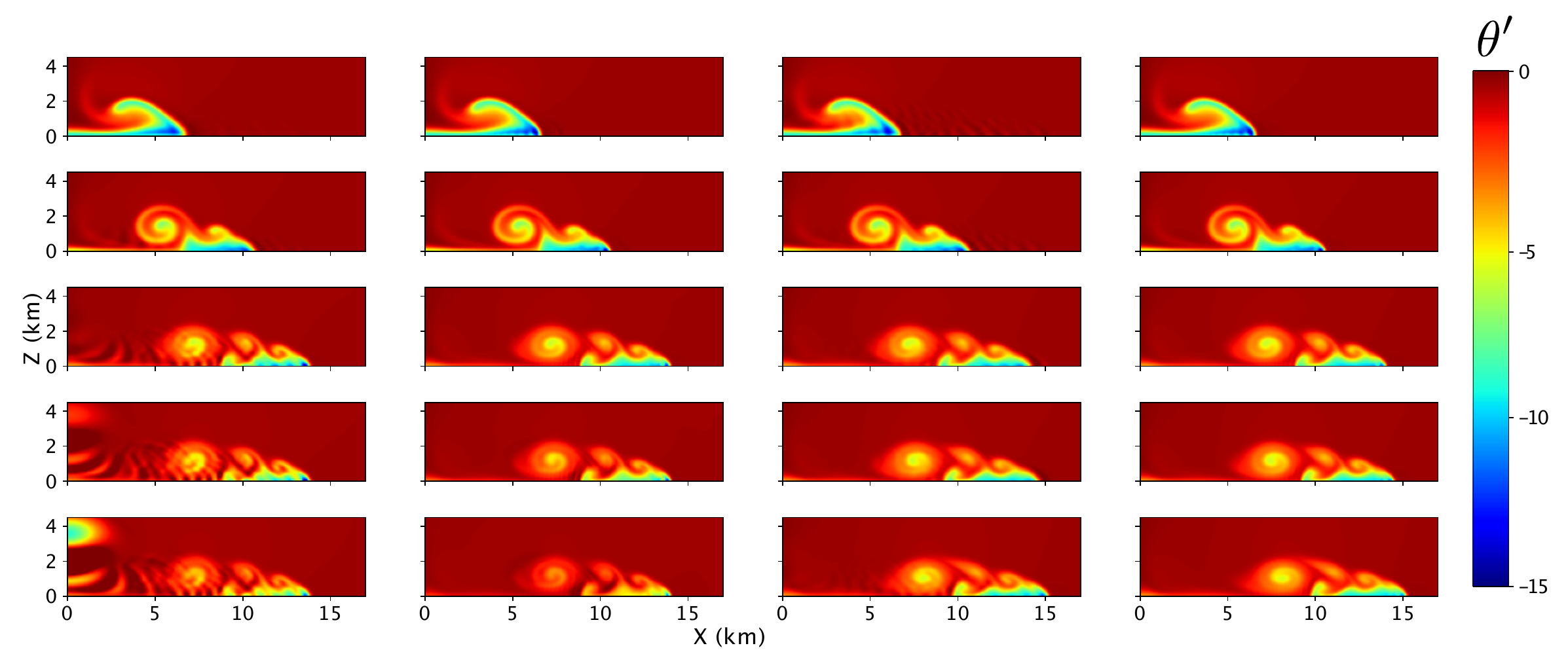}  
            \put(11,39){DMD}
            \put(33,39){HDMD}
            \put(56.5,39){PODI}
            \put(79,39){FOM}
            \put(11,36.5){\tiny{\textcolor{white}{$t$=400 s}}}
            \put(11,29.1){\tiny{\textcolor{white}{$t$=600 s}}}
            \put(11,21.8){\tiny{\textcolor{white}{$t$=820 s}}}
            \put(11,14.5){\tiny{\textcolor{white}{$t$=852 s}}}
            \put(11,7){\tiny{\textcolor{white}{$t$=900 s}}}

            \put(34,36.5){\tiny{\textcolor{white}{$t$=400 s}}}
            \put(34,29.1){\tiny{\textcolor{white}{$t$=600 s}}}
            \put(34,21.8){\tiny{\textcolor{white}{$t$=820 s}}}
            \put(34,14.5){\tiny{\textcolor{white}{$t$=852 s}}}
            \put(34,7){\tiny{\textcolor{white}{$t$=900 s}}}

            \put(56.5,36.5){\tiny{\textcolor{white}{$t$=400 s}}}
            \put(56.5,29.1){\tiny{\textcolor{white}{$t$=600 s}}}
            \put(56.5,21.8){\tiny{\textcolor{white}{$t$=820 s}}}
            \put(56.5,14.5){\tiny{\textcolor{white}{$t$=852 s}}}
            \put(56.5,7){\tiny{\textcolor{white}{$t$=900 s}}}

            \put(79,36.5){\tiny{\textcolor{white}{$t$=400 s}}}
            \put(79,29.1){\tiny{\textcolor{white}{$t$=600 s}}}
            \put(79,21.8){\tiny{\textcolor{white}{$t$=820 s}}}
            \put(79,14.5){\tiny{\textcolor{white}{$t$=852 s}}}
            \put(79,7){\tiny{\textcolor{white}{$t$=900 s}}}
            
      \end{overpic}
      \caption{Density current: comparison of evolution of $\theta'$ given by the ROMs (first 3 columns) and the FOM (last column) for $\delta = 0.99$.
      }
      
        \label{fig:POD_DMD_HDMD_Alltime_DC}
\end{figure}

In view of the results in Sec.~\ref{sec:RTB}, we set
the energy threshold $\delta$ to $99\%$. See Tab.~\ref{tab:DC_1st_Modes} for the number of modes retained for the different methods.  
\fig{fig:POD_DMD_HDMD_Alltime_DC}
compares the evolution of the potential temperature perturbation given by DMD, HDMD, and PODI with the evolution computed by the FOM. The top two rows in 
\fig{fig:POD_DMD_HDMD_Alltime_DC} correspond to solutions belonging to the training set. We see that 
all three methods can accurately identify the main flow structure, but the DMD and PODI solutions show some instability for $x < 5$ Km and $x \in [10,15]$ Km, respectively. These instabilities, that are especially evident at $t = 600$ s, are not present in the HDMD solutions.
The bottom three rows in \fig{fig:POD_DMD_HDMD_Alltime_DC} are not associated with the training set. 
We see that the instabilities in the DMD solution grow in time and expand to the majority of the domain, making the DMD prediction of the system dynamics off.
It appears that the DMD solution gives a large weight to many snapshots associated to previous times. Indeed, in the DMD solution for $t = 852$ s we can observe the time history of the system dynamics, i.e. the fall of the cold bubble and the front propagation. We note also that the more time passes, the larger the weights for the ``past'' snapshots become, i.e., the entire evolution of the flow structures becomes more visible. In fact, the DMD solution for $t = 900$ s clearly shows the initial cold bubble, which is not present in the DMD solution for $t = 820$ s. We remark that the DMD solutions for the rising bubble in \fig{fig:Vis_POD_DMD_HDMD_RTB_99} are affected by the same issue.
The HDMD algorithm fixes such issue and 
provides a very good prediction of $\theta'$ field for $t = 820$ s and $t = 852$ s. For $t = 900$ s, the HDMD solution compares less favorably with the FOM solution, although no instability emerges. Finally, the PODI solution suppresses the instability for $x \in [10,15]$ Km as time passes and is accurate for $t = 820, 852$ s. For $t = 900$ s though, instabilities arise in the PODI solution for $x < 5$ Km. 

\begin{figure}
    \centering \includegraphics[width=75mm,scale=0.5]{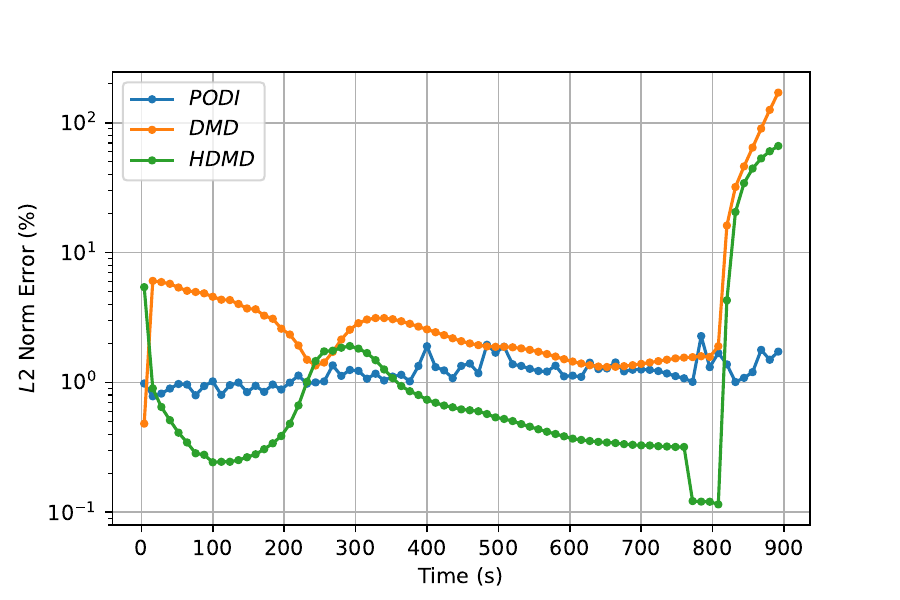}
        \caption{Density current: time evolution of the $L^2$ error \eqref{eq:l2Error} for DMD, HDMD, and PODI for $\delta= 99 \%$.}   
        \label{fig:Error_alltime_DC}
\end{figure}

For a more quantitative comparison, we show the time evolution of $L^2$ error  \eqref{eq:l2Error} in \fig{fig:Error_alltime_DC}.
Although qualitatively the HDMD solution compares well with the FOM solution (See \fig{fig:POD_DMD_HDMD_Alltime_DC}), we see that
both DMD and HDMD do a poor job in predicting the future behavior of the system in the $L^2$ norm. 
{Our suspicion that HDMD captures the
shape of the flow structures well, but not their propagating speed, is confirmed with this test: the large $L^2$ error for HDMD is mainly due to the fact that the front propagation slows down in comparison to the FOM solution. In fact, in the FOM solution the front is located around $x = 15$ Km at $t = 900$ s, while in the HDMD solution is around $x = 14$ Km at the same time.}
Their accuracy of DMD and HDMD is even worse than for the previous benchmark, as it is clear from comparing \fig{fig:Error_alltime_DC} to the bottom panel of \fig{fig:Error_in_time_POD_DMD_HDMD_RTB}. 
Specifically, the $L^2$ error  \eqref{eq:l2Error} for DMD increases from about $60\%$ to around $140 \%$, while the error for HDMD increases from $22\%$ to about $60\%$.
This is a further confirmation that it is more challenging to capture the flow dynamics in the density current test than the thermal rising bubble test. 
On the other hand, thanks to the interpolatory approach the PODI solution maintains a good accuracy throughout the entire time interval, oscillating around 1\% during fall of the bubble and reaching a maximum of about 3\% in the horizontal convection phase. 

To acknowledge the fact that this benchmark features mainly vertical dynamics before mainly horizontal dynamics, we now restrict the database to the computed solutions for $t \geq 280$ s, i.e., we discard 
the snapshots associated to the mainly vertical motion. This idea, which is intended to improve the results in \fig{fig:POD_DMD_HDMD_Alltime_DC} and \ref{fig:Error_alltime_DC}, is inspired from \cite{girfoglio2023hybrid, strazzullo2022consistency}.
The first snapshot of the new database coincides with the computed $\theta'$ field at $t = 280$ s shown in the right panel of \fig{fig:DC_InitialCon} and the total number of snapshots is 155.
We use the $85\%$ of the snapshots as training set, following time order for DMD and HDMD and randomly for PODI. 
We will refer to this subset of the database as the second training set.
The choice of taking $85\%$ of the database (i.e., 133 snapshots) is to have the same prediction time for DMD and HDMD as before, i.e., $t =[808,900]$ s.
This will make the comparison with the results in \fig{fig:POD_DMD_HDMD_Alltime_DC} and \ref{fig:Error_alltime_DC} fair  for DMD/HDMD. 

\begin{figure}
    \centering \includegraphics[width=80mm,scale=0.5]{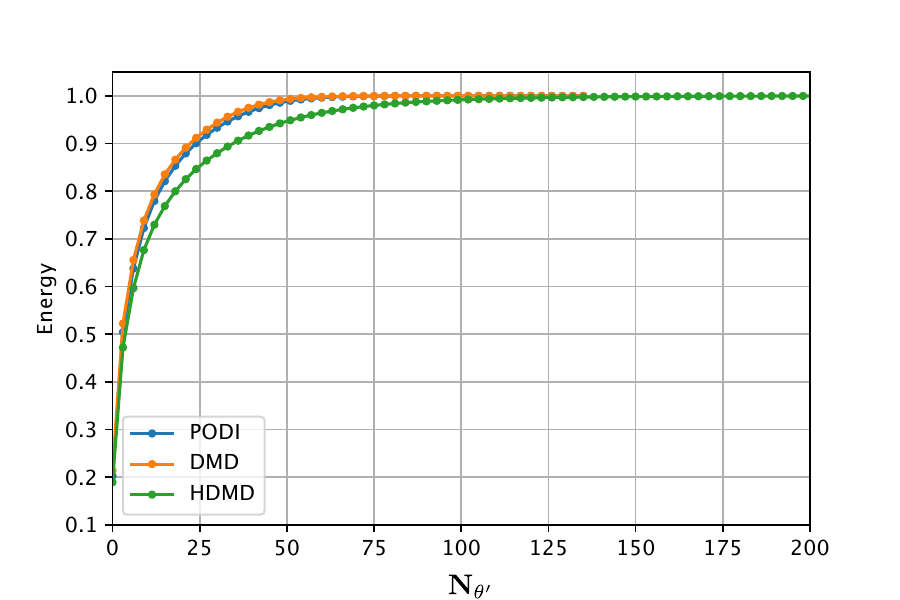}
        \caption{Density current, second training set: cumulative energy  $E$ in \eqref{k_equation} for the three ROM approaches under consideration. 
        }  
        \label{fig:modes_energy_theta_DC_2nd}
\end{figure}

\fig{fig:modes_energy_theta_DC_2nd} shows the cumulative energy  $E$ in \eqref{k_equation} for all the ROMs. We set again the energy threshold $\delta$ to 99\%, i.e., we retain 49 modes for DMD, 96 modes for HDMD and 52 modes for PODI. 
Since we are considering only the front propagation phase, the number of modes is less than in the bottom row  of Tab.~\ref{tab:DC_1st_Modes}.

\begin{figure}[t]
    \begin{subfigure}{\linewidth} 
    \centering
        \begin{overpic}[width=1.0\textwidth, grid=false]{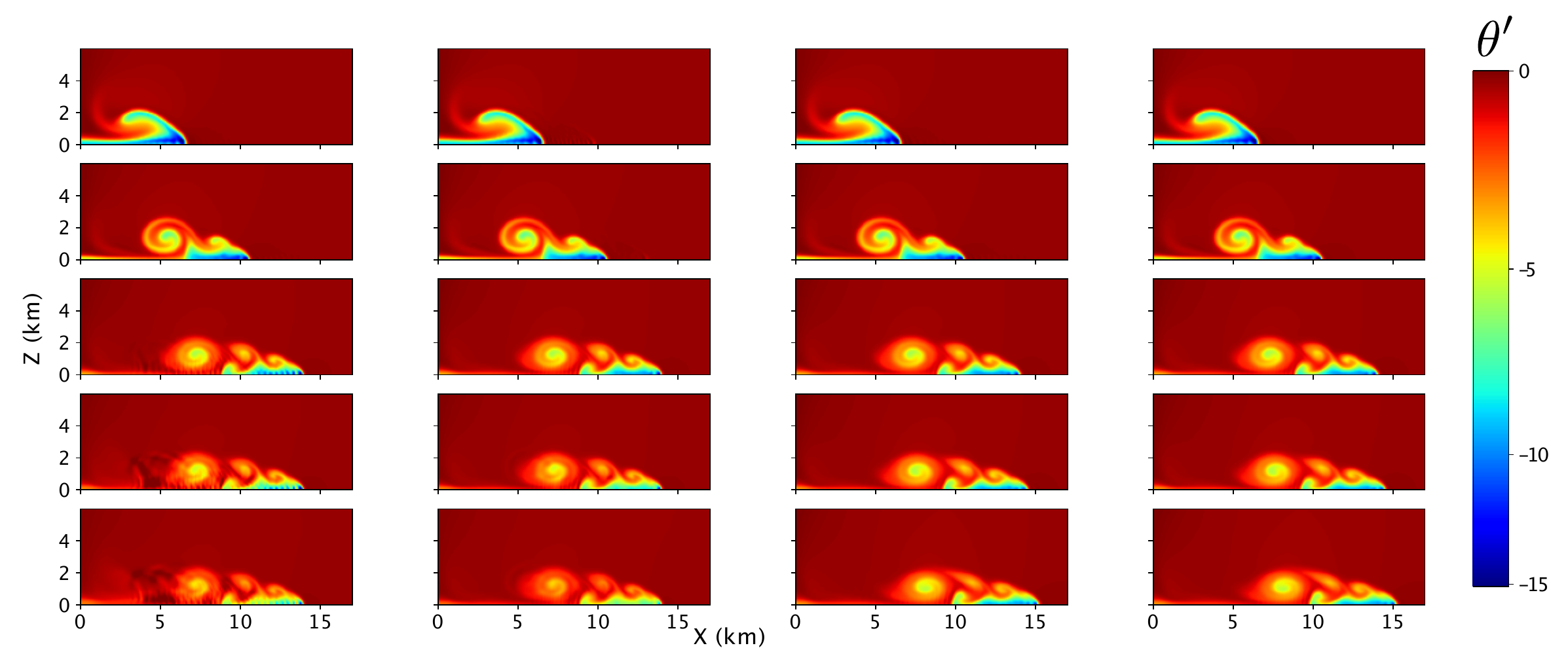}  
            \put(11,39.5){DMD}
            \put(33,39.5){HDMD}
            \put(56.5,39.5){PODI}
            \put(79,39.5){FOM}
            \put(11,37){\tiny{\textcolor{white}{$t$=400 s}}}
            \put(11,29.6){\tiny{\textcolor{white}{$t$=600 s}}}
            \put(11,22.3){\tiny{\textcolor{white}{$t$=820 s}}}
            \put(11,15){\tiny{\textcolor{white}{$t$=852 s}}}
            \put(11,7.5){\tiny{\textcolor{white}{$t$=900 s}}}

            \put(34,37){\tiny{\textcolor{white}{$t$=400 s}}}
            \put(34,29.6){\tiny{\textcolor{white}{$t$=600 s}}}
            \put(34,22.3){\tiny{\textcolor{white}{$t$=820 s}}}
            \put(34,15){\tiny{\textcolor{white}{$t$=852 s}}}
            \put(34,7.5){\tiny{\textcolor{white}{$t$=900 s}}}

            \put(56.5,37){\tiny{\textcolor{white}{$t$=400 s}}}
            \put(56.5,29.6){\tiny{\textcolor{white}{$t$=600 s}}}
            \put(56.5,22.3){\tiny{\textcolor{white}{$t$=820 s}}}
            \put(56.5,15){\tiny{\textcolor{white}{$t$=852 s}}}
            \put(56.5,7.5){\tiny{\textcolor{white}{$t$=900 s}}}

            \put(79,37){\tiny{\textcolor{white}{$t$=400 s}}}
            \put(79,29.6){\tiny{\textcolor{white}{$t$=600 s}}}
            \put(79,22.3){\tiny{\textcolor{white}{$t$=820 s}}}
            \put(79,15){\tiny{\textcolor{white}{$t$=852 s}}}
            \put(79,7.5){\tiny{\textcolor{white}{$t$=900 s}}}
        \end{overpic}
    \end{subfigure}  
\caption{ 
Density current, second training set: comparison of evolution of $\theta'$ given by the ROMs (first 3 columns) and the FOM (last column) for $\delta = 0.99$.} 
    \label{fig:POD_DMD_HDMD_DC_2ndInCond}
\end{figure}
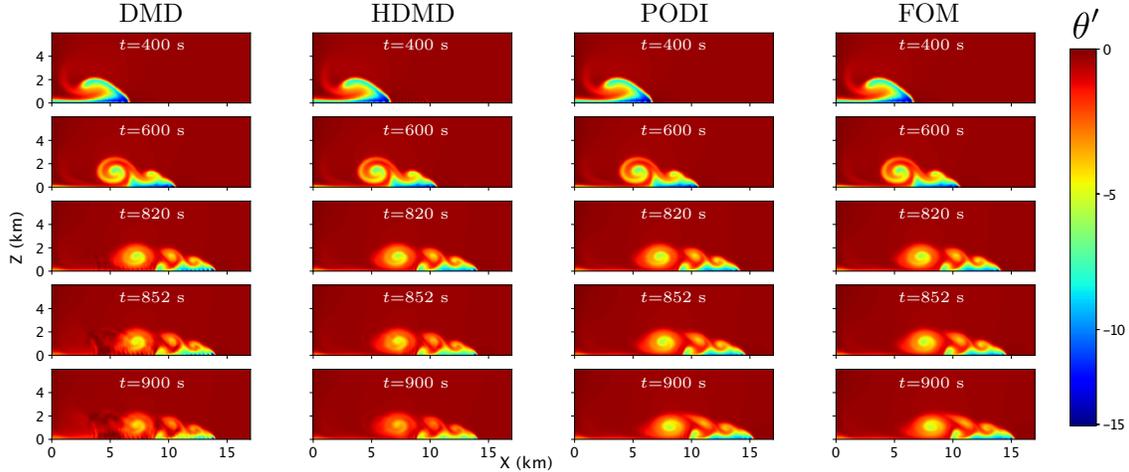


Now that the basis functions have been identified, we proceed with a qualitative comparison reported in  \fig{fig:POD_DMD_HDMD_DC_2ndInCond}.  
Just like in \fig{fig:POD_DMD_HDMD_Alltime_DC}, the top two rows in \fig{fig:POD_DMD_HDMD_DC_2ndInCond} correspond to solutions belonging to the training set. We see an improved system identification for DMD and PODI, with none of the instabilities observed in the first two rows of \fig{fig:POD_DMD_HDMD_Alltime_DC}.
By looking at the bottom three rows in \fig{fig:POD_DMD_HDMD_DC_2ndInCond}, we observe an improvement also in the system prediction, especially in the case of PODI whose solutions are 
oscillations-free and very similar to the FOM solutions. The DMD solutions, although improved with respect to \fig{fig:POD_DMD_HDMD_Alltime_DC}, are still affected by instabilities for $x < 5$ Km that especially evident for $t = 820$ s and $t = 900$ s. 

\begin{figure}
\centering
\begin{overpic}[width=0.48\textwidth, grid=false]{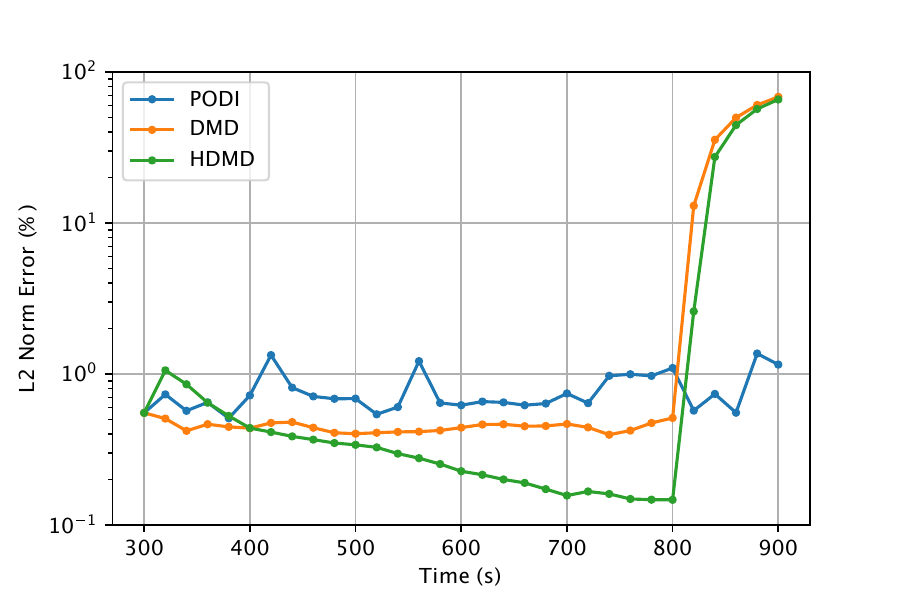} 
\end{overpic} 
\caption{Density current, second training set: time evolution of the $L^2$ error \eqref{eq:l2Error} for DMD, HDMD, and PODI for energy threshold $\delta= 0.99 \%$.} 
\label{fig:Error_in_time_POD_DMD_HDMD_DC_2nd}
\end{figure}

Finally, let us take a look at the time evolution of $L^2$ error \eqref{eq:l2Error} in \fig{fig:Error_in_time_POD_DMD_HDMD_DC_2nd}. The improvement in accuracy is clearly observable by comparing \fig{fig:Error_in_time_POD_DMD_HDMD_DC_2nd} with \fig{fig:Error_alltime_DC}. 
The DMD algorithm shows a significant improvement, with its performance getting very close to the HDMD. Indeed, the maximum error
for DMD decreases from about $120 \%$ to about $70 \%$, which is still unacceptable though.
The error for PODI remains around 1\% throughout the entire time interval. 
We remark that PODI showed comparable accuracy for the rising thermal bubble when $\delta = 0.99$ (see \fig{fig:Error_in_time_POD_DMD_HDMD_RTB}, bottom panel).

\subsubsection{Parametric reconstruction}
\label{Parametric_rec_DC}

As mentioned in Sec.~\ref{sec:ROM}, time is the parameter the DMD and HDMD methods were designed to handle, while PODI can be used for computational studies involving physical parameters too. Hence, in this subsection we investigate the accuracy of PODI in a parametric study involving parameter $\theta_s$ in \eqref{eq:initila_pot_temp_DC}.

Let $\theta_s \in [5, 10]$ K. Like in the case of time, 
we choose a uniform sample distribution for $\theta_s$ with 11 sampling points. For each of these 11 values of $\theta_s$, we run a simulation for the entire time interval of interest, i.e., $[0, 900]$ s and collect snapshots every $4$ s. The total number of snapshots in the database is
2475. Once again, we randomly select 90\% 
of them (i.e., 2227) for the training set. The remaining snapshots are used for validation. 

\begin{figure}
    \centering \includegraphics[width=80mm,scale=0.5]{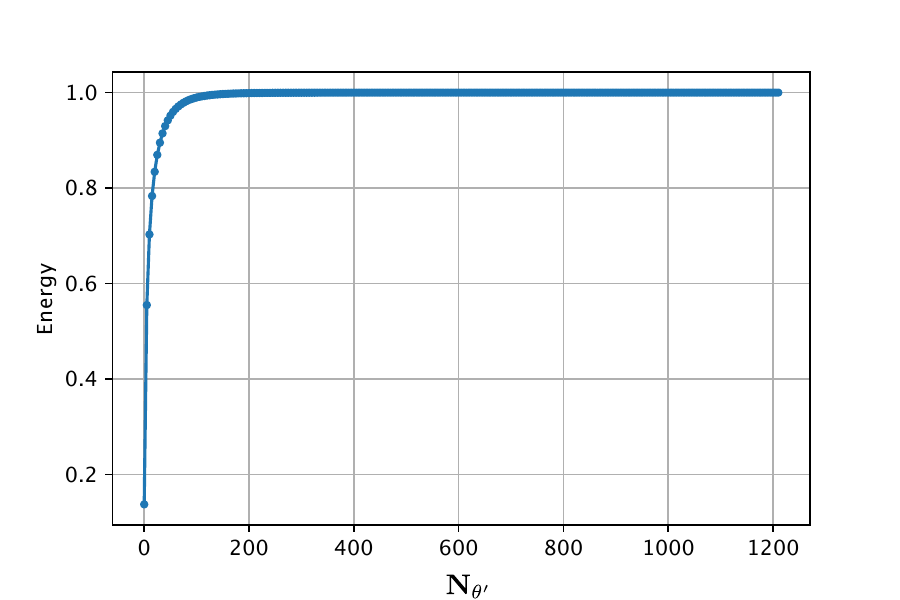}
        \caption{Density current, parametric study: cumulative energy  $E$ in \eqref{k_equation} as $N_{\theta'}$ varies.}    \label{fig:modes_energy_theta_paramtetic}
\end{figure}

\fig{fig:modes_energy_theta_paramtetic} shows cumulative energy  $E$ in \eqref{k_equation} as $N_{\theta'}$ varies. We truncated the graph at $N_{\theta'} = 1200$ since $E$ increases very steeply and gets over 99\% for rather small values of $N_{\theta'}$. Indeed, to retain 99\% of the energy we only need 120 modes.
 
\begin{figure}
 \vspace{1cm}
    \begin{subfigure}{\linewidth} 
        \centering
        \begin{overpic}[width=0.8\textwidth, grid=false]{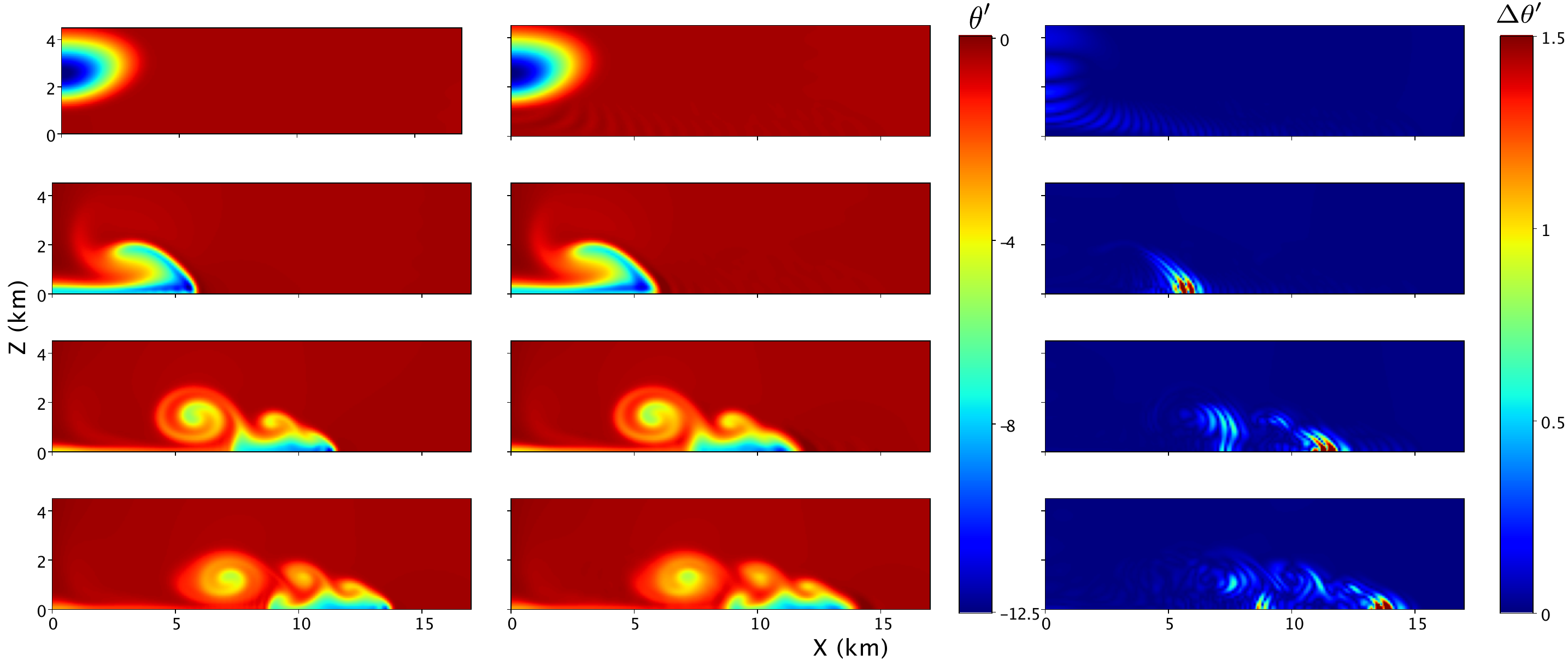} 
            \put(40,47){$\theta_s = 6.25$}
            \put(14,42){FOM}
            \put(43,42){PODI}
            \put(79,42){$\Delta {\theta} ^\prime$ }
            \put(14,39){\tiny{\textcolor{white}{$t$=84 s}}}
            \put(14,29){\tiny{\textcolor{white}{$t$=404 s}}}
            \put(14,18.8){\tiny{\textcolor{white}{$t$=724 s}}}
            \put(14,8.7){\tiny{\textcolor{white}{$t$=884 s}}}

            \put(43,39){\tiny{\textcolor{white}{$t$=84 s}}}
            \put(43,29){\tiny{\textcolor{white}{$t$=404 s}}}
            \put(43,18.8){\tiny{\textcolor{white}{$t$=724 s}}}
            \put(43,8.7){\tiny{\textcolor{white}{$t$=884 s}}}

            \put(78,39){\tiny{\textcolor{white}{$t$=84 s}}}
            \put(78,29){\tiny{\textcolor{white}{$t$=404 s}}}
            \put(78,18.8){\tiny{\textcolor{white}{$t$=724 s}}}
            \put(78,8.7){\tiny{\textcolor{white}{$t$=884 s}}}

       \end{overpic}
    \end{subfigure}
   \par\bigskip
   \vspace{1cm}
    \begin{subfigure}{\linewidth} 
    \centering
        \begin{overpic}[width=0.8\textwidth, grid=false]{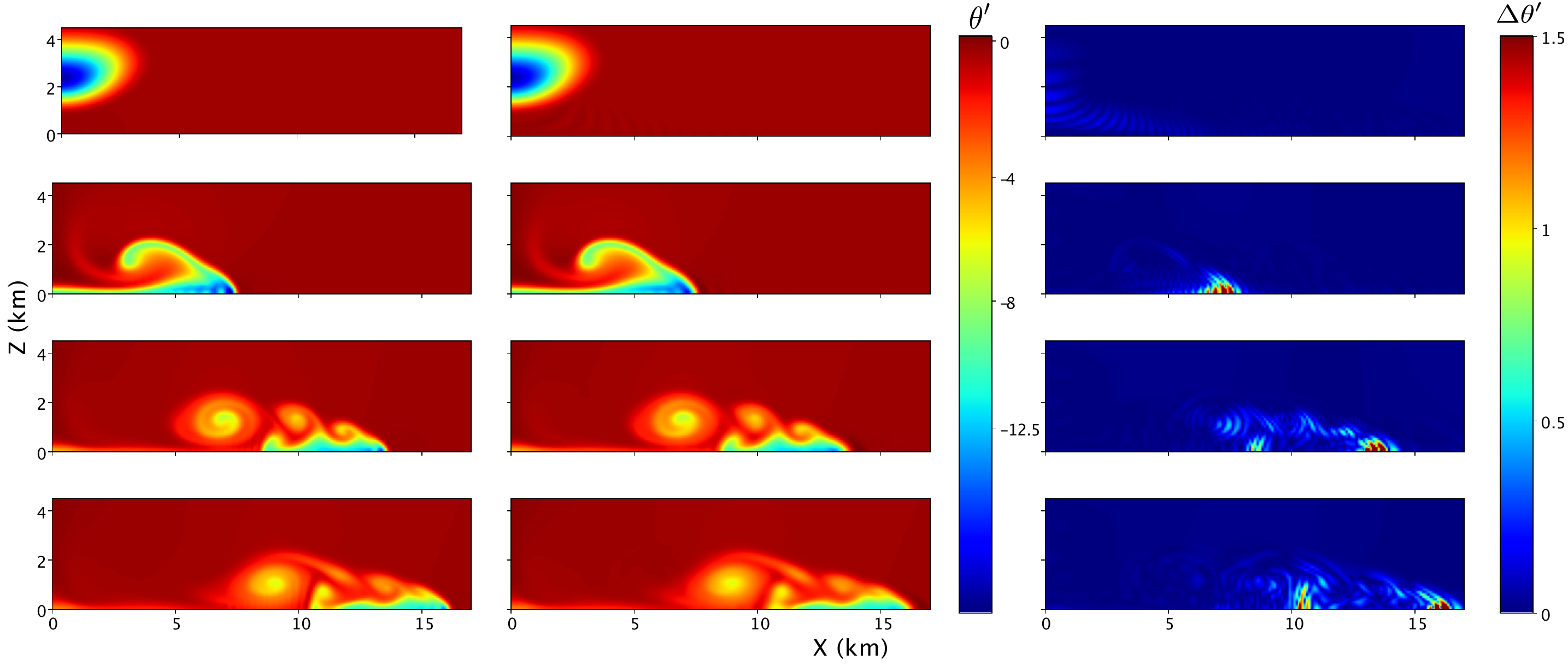}  
            \put(40,47){$\theta_s = 8.75$}
            \put(14,42){FOM}
            \put(43,42){PODI}
            \put(79,42){$\Delta {\theta} ^\prime$ }
            \put(14,39){\tiny{\textcolor{white}{$t$=84 s}}}
            \put(14,29){\tiny{\textcolor{white}{$t$=404 s}}}
            \put(14,18.8){\tiny{\textcolor{white}{$t$=724 s}}}
            \put(14,8.7){\tiny{\textcolor{white}{$t$=884 s}}}

            \put(43,39){\tiny{\textcolor{white}{$t$=84 s}}}
            \put(43,29){\tiny{\textcolor{white}{$t$=404 s}}}
            \put(43,18.8){\tiny{\textcolor{white}{$t$=724 s}}}
            \put(43,8.7){\tiny{\textcolor{white}{$t$=884 s}}}

            \put(78,39){\tiny{\textcolor{white}{$t$=84 s}}}
            \put(78,29){\tiny{\textcolor{white}{$t$=404 s}}}
            \put(78,18.8){\tiny{\textcolor{white}{$t$=724 s}}}
            \put(78,8.7){\tiny{\textcolor{white}{$t$=884 s}}}
        \end{overpic}
    \end{subfigure}
\caption{Density current, parametric study: FOM solutions (left column), PODI solutions (center column), and difference between the two in absolute value (third column)  for $\theta_s=6.25$ 
(top panel) and $\theta_s=8.75$ in (bottom panel).}
    \label{fig:Vis_Parametric_Rom_DC}
\end{figure}
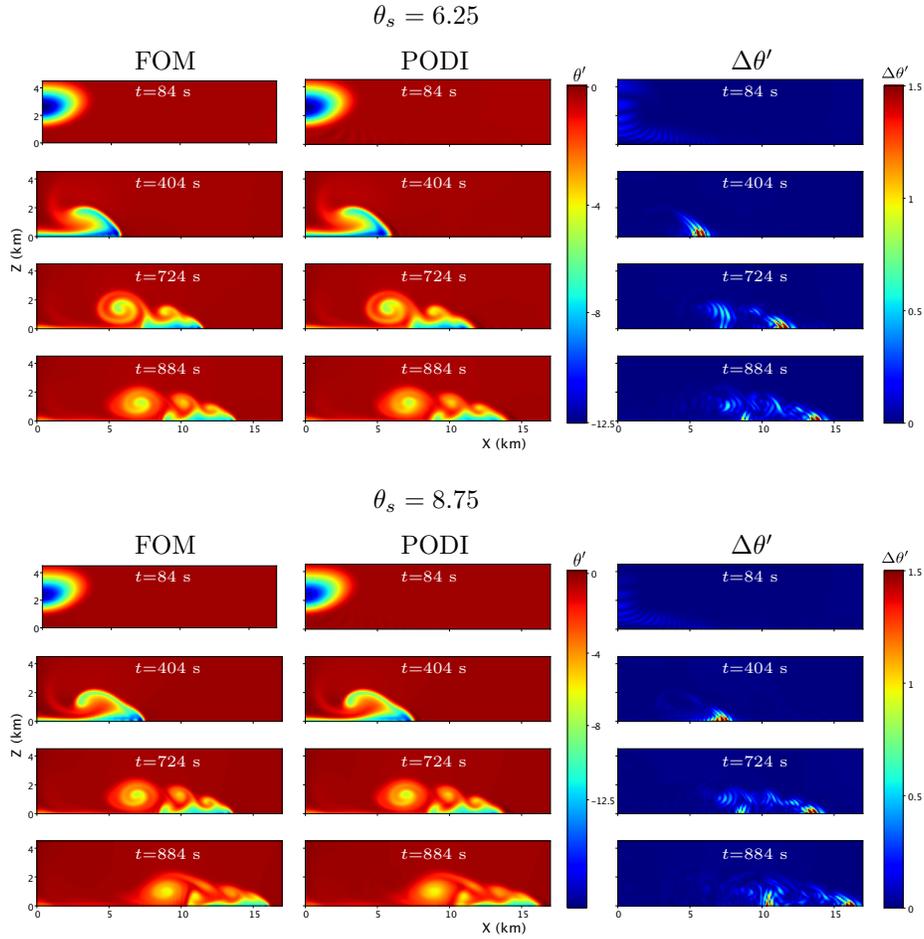

To evaluate the accuracy of PODI in the parametric study, we consider two values that are not associated to snapshots in the training set: $\theta_s=6.25$ and $\theta_s=8.75$. 
A qualitative comparison between the PODI and FOM solutions is shown in \fig{fig:Vis_Parametric_Rom_DC}, together with the difference between the two in absolute value.
We see that the difference in the FOM and PODI solutions, which does not exceed 1.5 K in absolute value,
is localized and confined to the region of the main flow structures. We also observe that the difference in absolute value is smaller during the vertical fall of the bubble (top row in both panels) than during the front propagation (bottom three rows in both panels).



\begin{figure}
\centering
\begin{overpic}[width=0.49\textwidth, grid=false]{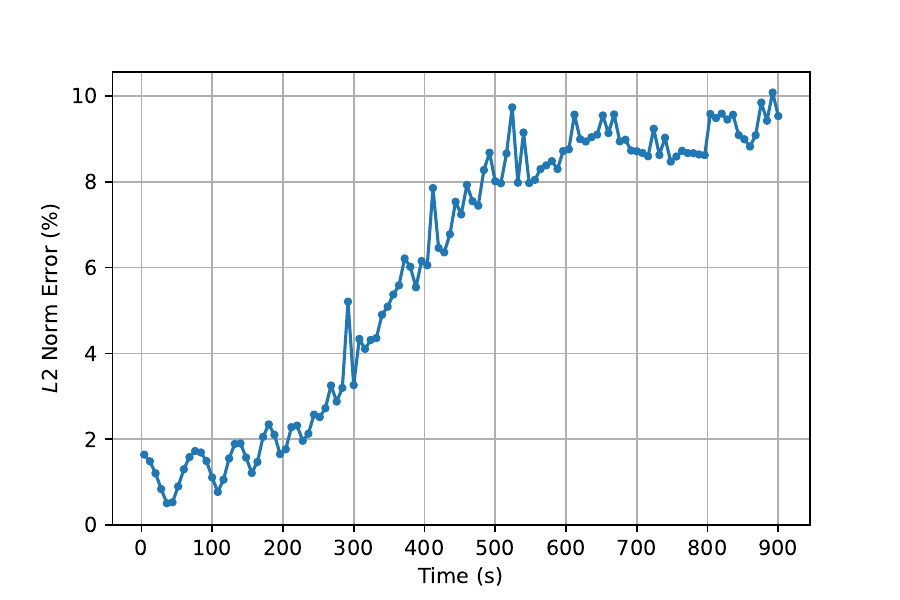}  
            \put(45,61){$\theta_s = 6.25$}
\end{overpic}            
\begin{overpic}[width=0.49\textwidth, grid=false]{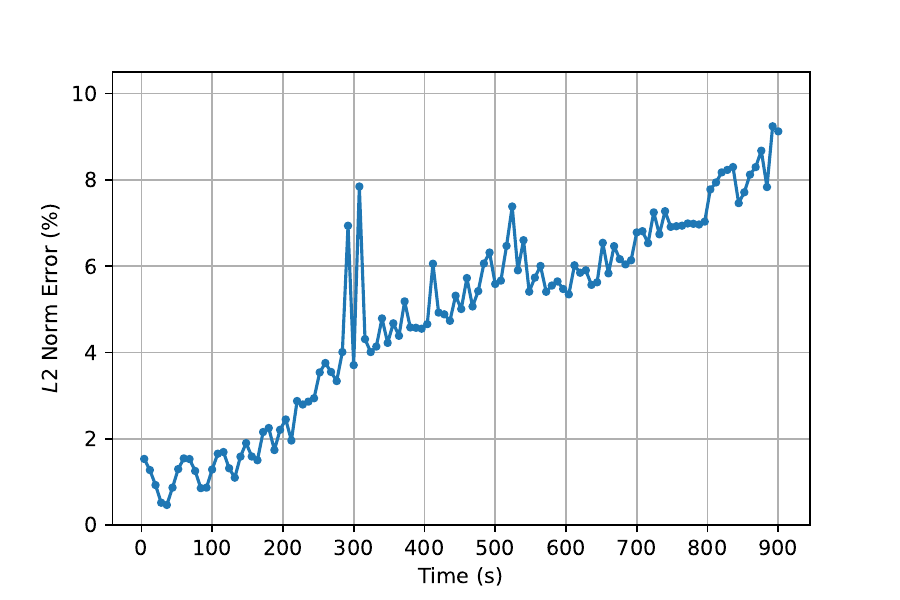}  
            \put(45,61){$\theta_s = 8.75$}
\end{overpic}
\caption{Density current, parametric study:  evolution of the $L^2$ error \eqref{eq:l2Error} for 
$\theta_s = 6.25$ (left) and $\theta_s = 8.75$ (right).}
\label{fig:Error_parametric}
\end{figure}

The time evolution of $L^2$ error \eqref{eq:l2Error} for $\theta_s = 6.25$ and $\theta_s = 8.75$ reported in \fig{fig:Error_parametric}
confirms that the error is smaller initially (i.e., during the fall of the bubble) and increases at later time instances (i.e., when during the front propagation). We remark that for both values of $\theta_s$ the error is lower than 10\% for the entire time interval. This accuracy can be improved upon by discarding the snapshots associated to the fall of the bubble from the database, as shown in Sec.~\ref{sec:Time_rec_DC}.



\section{Concluding remarks}\label{sec:conc}
With the goal of reducing the computational time to forecast regional atmospheric flow, we considered three data driven reduced order modeling techniques: a ROM specifically designed for system prediction called DMD, an improvement of DMD called HDMD, and an interpolatory ROM called PODI. PODI has the advantage over DMD and HDMD to allow for parametric studies, i.e., it can handle physical parameters in the same way it handles time. We applied the three ROMs to two well-known benchmarks for mesoscale flow and compared their accuracy in system identification and prediction of the system behavior, and in terms of computational time. Since the use of ROMs for the prediction of atmospheric flow is still in its infancy, our work in this paper has a few distinguishing elements:
(i) the ROMs are applied to the simulation of mesoscale flow, which features higher resolution than the simulation of global circulation; (ii) the ROMs are used for both system identification and prediction; and (iii) one ROM is used for a parametric study.

In the case where time is the only parameter of interest, our results show that all three ROMs are accurate in the identification of the system dynamics, although local instabilities are seen in the DMD and PODI solutions. The price to pay for the lack of oscillations and increased accuracy in the HDMD solutions is a substantial increase in computational time: the time of a HDMD simulation is two orders of magnitude larger than the time of a DMD or PODI simulation. Although DMD and HDMD are intended for forecasts, the accuracy in the prediction of the system dynamics is low even when 99\% of the eigenvalue energy is retained and the snapshots in the database are tailored to the problem at hand. Thanks to the interpolatory approach, PODI maintains a good level of accuracy during the entire time interval of interest. This is true also when a physical parameter is varied within a parametric study. 


We believe that the results presented in this paper can be improved upon by using Machine Learning-based techniques that can better detect and reproduce the nonlinear behavior exhibited by the full order model. In particular, Convolutional Autoencoders and Long-Short Time Memory could improve
both the accuracy and efficiency of the methods in this paper \cite{gonzalez2018deep,maulik2021reduced, mohan2019compressed, shi2015convolutional}. 

\section{acknowledgments}\label{sec:akw}

We acknowledge the support provided by the European Research Council Executive Agency by the Consolidator Grant project AROMA-CFD "Advanced Reduced Order Methods with Applications in Computational Fluid Dynamics" - GA 681447, H2020-ERC CoG 2015 AROMA-CFD, the European Union’s Horizon 2020 research and innovation program under the Marie Skłodowska-Curie Actions, grant agreement 872442 (ARIA), PON “Research and Innovation on Green related
issues” FSE REACT-EU 2021 project, PRIN NA FROM-PDEs project, European High-Performance Computing Joint Undertaking project Eflows4HPC GA N. 955558 and INdAM-GNCS 2019-2021 projects. 
This work was also partially supported by US National Science Foundation through grant DMS-1953535 (PI A.~Quaini). 

\bibliographystyle{unsrt}
\bibliography{mybib}

\end{document}